\documentclass[reqno,12pt]{amsart}

\usepackage{amsmath,amssymb,amsthm,amsfonts,amsxtra}
\usepackage{fullpage}

\begin{document}

\newtheorem{thm}{Theorem}
\newtheorem{cor}{Corollary}
\newtheorem{lem}{Lemma}
\newtheorem{prop}{Proposition}

\def\Ref#1{Ref.~\cite{#1}}

\def\const{\text{const.}}
\def\Rnum{{\mathbb R}}
\def\sgn{{\rm sgn\;}}
\def\arctanh{\mathrm{arctanh}}

\def\X{\mathrm{X}}
\def\pr{\mathrm{pr}}
\def\Esp{{\mathcal E}}

\def\tilu{\tilde u}
\def\tilg{\tilde g}
\def\tilG{\tilde G}

\def\dens{T}
\def\flux{\Phi}
\def\triv{\Theta}

\tolerance=10000
\allowdisplaybreaks[4]

\title{Conservation laws and variational structure\\\ of damped nonlinear wave equations}

\author{
Stephen C. Anco$^1$,\\
Almudena P. M\'arquez$^2$,
Tamara M. Garrido$^2$,
Mar\'ia L. Gandarias$^2$
\\\\
${}^1$D\lowercase{\scshape{epartment}} \lowercase{\scshape{of}} M\lowercase{\scshape{athematics and}} S\lowercase{\scshape{tatistics}}\\
B\lowercase{\scshape{rock}} U\lowercase{\scshape{niversity}}\\
S\lowercase{\scshape{t.}} C\lowercase{\scshape{atharines}}, ON L2S3A1, C\lowercase{\scshape{anada}} \\
\\
${}^2$D\lowercase{\scshape{epartment}} \lowercase{\scshape{of}} M\lowercase{\scshape{athematics}}\\
U\lowercase{\scshape{niversity of}} C\lowercase{\scshape{adiz}}\\
11510 P\lowercase{\scshape{uerto}} R\lowercase{\scshape{eal}}, C\lowercase{\scshape{adiz}}, S\lowercase{\scshape{pain}}\\
}


\begin{abstract}
All low-order conservation laws are found
for a general class of nonlinear wave equations in one dimension 
with linear damping which is allowed to be time-dependent. 
Such equations arise in numerous physical applications
and have attracted much attention in analysis. 
The conservation laws describe 
generalized momentum and boost momentum, conformal momentum, 
generalized energy, dilational energy, and light-cone energies. 
Both the conformal momentum and dilational energy have no counterparts 
for nonlinear undamped wave equations in one dimension. 
All of the conservation laws are obtainable through Noether's theorem, 
which is applicable because the damping term can be transformed into 
a time-dependent self-interaction term by a change of dependent variable. 
For several of the conservation laws, the corresponding variational symmetries 
have a novel form which is different than any of the well known variation symmetries
admitted by nonlinear undamped wave equations in one dimension. 
\end{abstract}

\maketitle

\section{Introduction}

The wave equation $u_{tt} =c^2 u_{xx}$ is a commonplace simple model 
for wave propagation and vibrations in one dimension
(see e.g. \Ref{Whitham-book}), 
where $c$ is the wave speed and $u(x,t)$ is the wave amplitude.
In real world applications,
it is of interest to consider more general models 
that include the effects of linear damping and self-interaction:
\begin{equation}\label{waveeqn}
u_{tt} + a(t) u_t + g(u) =c^2 u_{xx} 
\end{equation}
where $a(t)>0$ is the damping coefficient and $g(u)\neq0$ is the self-interaction term.  
Damped wave equations of this type occur in numerous physical applications. 

For instance, in the situation when there is no self-interaction, $g''(u)\equiv 0$, 
and the damping coefficient is constant, $a'(t)\equiv 0$, 
the wave equation becomes the telegrapher's equation 
$u_{tt} + a u_t + g_0 u =c^2 u_{xx}$, $g_0=\const$, 
after a shift in $u$. 
For $g_0=0$, this is the classical damped linear wave equation,
which describes frictionally damped mechanical vibrations \cite{Wal}. 
When $g_0\neq 0$, 
the telegrapher's equation describes electrical transmission lines 
as well as other electromagnetic signal applications \cite{Kra},
and heat flow with a time-relaxation conduction law \cite{JosPre,Cha}. 

In the more interesting situation when nonlinear self-interaction is relevant, 
examples of physical models based on the nonlinear equation \eqref{waveeqn} 
with $g''(u)\not\equiv 0$ include: 
a special case of a general nonlinear transmission line system \cite{Sco}
in which the inductance and resistance are functions of the voltage; 
an approximation for reaction-diffusion systems with transport memory 
which arises in the physics of materials \cite{ManHurKen}
and in the dynamics of epidemics and population ecology \cite{KenGiu}; 
a theory of hyperbolic nonlinear heat conduction \cite{JorLam}. 
Analysis of the global behaviour of solutions has also attracted a lot of attention 
for both time-independent damping $a'(t)\equiv 0$ 
\cite{NakOno,LiZho,TodYor,IkeOht} 
and time-dependent damping $a'(t)\not\equiv 0$ 
\cite{Nis,LinNisZha,IkeSobWak,IkeWak}, 
especially for the case when $g(u) = k u|u|^p$ is a power nonlinearity, with $p>0$. 

To understand an important implication of damping, 
it is useful to consider the energy 
$E=\int_\Rnum \big( \tfrac{1}{2}(u_t{}^2 + c^2 u_x{}^2) + \smallint g(u)\, du \big)\, dx$
which is conserved for undamped nonlinear wave equations 
\begin{equation}\label{waveeqn.undamped}
u_{tt} + g(u) =c^2 u_{xx} . 
\end{equation}
Energy is no longer conserved when damping is present, 
since $a(t)>0$ shows that $\frac{d}{dt}E =-a(t)\int_\Rnum u_t{}^2\, dx<0$
for all non-equilibrium solutions, $u_t\not\equiv 0$, 
of the damped wave equation \eqref{waveeqn}. 
As a consequence, from the viewpoint of conserved quantities,
damped wave equations could be expected to lack 
any energy-type conservation laws.

The main purpose of the present paper is to show that, on the contrary, 
an interesting variety of generalized energy-momentum type conservation laws 
do exist for certain forms of self-interaction and damping, 
despite the fact that the ordinary energy is decreasing for non-equilibrium solutions.
These conservation laws yield conserved quantities describing
generalized momentum and boost momentum, conformal momentum, 
generalized energy, dilational energy, and light-cone energies. 
Of particular interest is the fact that there are no counterparts of 
the dilation energy and the conformal momentum 
for nonlinear undamped wave equations \eqref{waveeqn.undamped}, 
although such conserved quantities exist in all higher dimensions 
when $g(u)$ is a certain dimension-dependent power nonlinearity. 

In one sense, the existence of these new conserved quantities 
can be understood from the observation that 
the damped nonlinear wave equation \eqref{waveeqn} has a variational structure. 
In particular, this wave equation becomes an Euler-Lagrange equation 
when it is multiplied by the factor $e^{2\int a(t)\,dt}$. 
Noether's theorem then provides a one-to-one correspondence between 
variational symmetries and conservation laws
as shown by general results in \Ref{Olv-book,BCA-book}.

However, in a deeper sense, the question is how to explain the existence of 
the variational symmetries that correspond to the new conserved quantities. 
Specifically, the symmetries underlying 
the conformal momentum and the light-cone energies 
respectively consist of a conformal transformation and translations in terms of 
exponentials involving light-cone coordinates. 
These symmetries are non-obvious and, 
at least for nonlinear wave equations with damping, 
are quite unexpected. 

Another surprising aspect is that the generalized energy and boost momentum 
arise when the damping has a special form such that it can be removed by 
a change of dependent variable, where both the original damped wave equation 
and the resulting undamped wave equation have the same type of nonlinearity. 
This describes a non-trivial equivalence transformation within 
the class of nonlinear wave equations \eqref{waveeqn}. 

For the purpose of computationally finding conservation laws, 
it is simplest to use the general method of multipliers 
\cite{Olv-book,AncBlu2002a,AncBlu2002b}, 
since there turns out to be a general direct connection between 
classifying all conserved quantities of a given form 
and classifying all multipliers of an associated form. 
As an example, all conserved quantities of (generalized) energy-momentum type 
correspond to multipliers that are linear, first-order in derivatives of $u$. 
The classifications that will be carried out here are for 
more general conserved quantities of first-order in derivatives of $u$,
which correspond to multipliers that are first-order but not necessarily linear. 
This is equivalent to classifying variational contact symmetries, 
including point symmetries as a special case. 

In Section~\ref{sec:prelims}, 
Noether's theorem and the connection between variational symmetries and multipliers 
for damped wave equations is explained. 

In Section~\ref{sec:equiv}, 
the conditions under which the damping term can be removed by a change of variable
are identified and solved. 
All low-order conservation laws and corresponding variational symmetries of 
the resulting nonlinear wave equation are presented. 

In Section~\ref{sec:results}, 
for damped nonlinear wave equations that are not transformable to an undamped wave equation, 
all low-order conservation laws are derived and the physical meaning of 
the resulting conserved quantities is discussed. 
In particular, a comparison is made with the well-known conservation laws of 
nonlinear undamped wave equations (Klein-Gordon equations). 

In Section~\ref{sec:symms},
the variational symmetry transformations associated to each conserved quantity 
are presented and their features are discussed. 

Some concluding remarks are made in Section~\ref{sec:conclude}.
An Appendix contains some remarks on the computations. 

Throughout, 
the mathematical setting is calculus on jet spaces \cite{Olv-book,Anc-review}. 
A relevant general treatment of 
symmetries, conservation laws, multipliers and variational symmetries for nonlinear PDEs 
can be found in \Ref{Olv-book,BCA-book,Anc-review}.

\section{Multipliers, variational structure, and Noether's theorem}\label{sec:prelims}

For generality in the subsequent exposition, 
damped wave equations in which the self-interaction is allowed to depend on $t$ and $x$ 
will be considered: 
\begin{equation}\label{waveeqn.gen}
u_{tt} + a(t) u_t + g(t,x,u) =c^2 u_{xx} . 
\end{equation}
This class of wave equations possesses a variational structure 
which arises through removing the damping term by a change of variables 
\begin{equation}\label{newvar}
v= e^{A}u
\end{equation}
where 
\begin{equation}\label{A}
A(t)=\tfrac{1}{2}\int^{t}_{t_0} a(t)\,dt, 
\quad
t_0=\const .
\end{equation}
The new variable $v$ satisfies an undamped wave equation 
\begin{equation}\label{new.waveeqn}
v_{tt} +h(t,x,v) =c^2 v_{xx}
\end{equation}
whose self-interaction is given by 
\begin{equation}\label{new.selfinteraction}
h(t,x,v)= -(\tfrac{1}{2}a_t + \tfrac{1}{4}a^2)v + e^{A} g(t,x,e^{-A}v) .
\end{equation}
Note that $h$ depends on $x$ if and only if $g$ depends on $x$. 
Also note that, in general, $h$ will depend on $t$, even if $g$ does not. 
The special situation when both $h$ and $g$ do not depend on $t$ will be addressed
in Section~\ref{sec:equiv}. 

The undamped wave equation \eqref{new.waveeqn} 
is well known to be an Euler-Lagrange equation 
\begin{equation}
\delta L^v/\delta v = v_{tt} +h(t,x,v) -c^2 v_{xx} =0
\end{equation} 
given by the Lagrangian 
$L^v = {-}\tfrac{1}{2} v_t{}^2 +\tfrac{1}{2}c^2 v_x{}^2 +H(t,x,v)$
where $H= \smallint h\,dv$. 
This implies that the original class of wave equations \eqref{waveeqn.gen}
also has a Lagrangian formulation. 
It is obtained directly by substitution of the change of variable \eqref{newvar} 
into $L^v$ to get the Euler-Lagrange equation 
\begin{equation}\label{ELeqn.gen}
\delta L^u/\delta u = \big( u_{tt} + a(t) u_t + g(t,x,u) -c^2 u_{xx} \big)e^{2A} =0
\end{equation} 
where the transformed Lagrangian is given by 
\begin{equation}\label{Lagr.gen}
L^u = \big( {-}\tfrac{1}{2} u_t{}^2 +\tfrac{1}{2} c^2 u_x{}^2 +G(t,x,u) \big)e^{2A}
\end{equation}
with $G=\smallint g\,du$. 
Note this Lagrangian can be changed by the addition of any total space-time divergence,
since such expressions are annihilated by the variational derivative. 

In the Euler-Lagrange equation \eqref{ELeqn.gen}, 
$e^{2A}$ is called a variational integrating factor. 
It connects variational symmetries to multipliers,
as will now be explained.

\subsection{Conservation laws and multipliers}

For a given wave equation \eqref{waveeqn.gen},
a {\em multiplier} is a function, $Q$, of $t$, $x$, $u$, and derivatives of $u$
such that it is non-singular on solutions $u(x,t)$ 
and its product with the wave equation is a total space-time divergence 
\begin{equation}\label{multr.eqn}
\big(u_{tt} + a(t) u_t + g(t,x,u) -c^2 u_{xx} \big)Q= D_t \dens + D_x \flux
\end{equation}
where $T$ and $\flux$ are some functions of $t$, $x$, $u$, and derivatives of $u$. 
Here $D_t$ and $D_x$ denote total derivatives.
This yields a local conservation law
\begin{equation}\label{conslaw}
(D_t \dens + D_x \flux)|_\Esp =0
\end{equation}
holding on the space $\Esp$ of all solutions $u(x,t)$ of the wave equation, 
where $T$ is a {\em conserved density} and $\flux$ is a {\em spatial flux}. 
The pair $(\dens,\flux)$ is called a {\em conserved current}. 

Integration of the conservation law \eqref{conslaw} 
over any spatial domain $\Omega \subseteq\Rnum$
gives a balance equation
\begin{equation}
\frac{d}{dt} \int_{\Omega} \dens\,dx = - \flux\Big|_{\partial\Omega}
\end{equation}
showing that the rate of change of the quantity
\begin{equation}\label{cons.integral}
C = \int_{\Omega} \dens\,dx\big|_\Esp
\end{equation}
is equal to the net flux through the boundary $\partial\Omega$ of $\Omega$. 
Hence, $C$ is a conserved quantity for solutions $u(x,t)$. 
It will be a constant of motion, namely, $\frac{d}{dt}C =0$, 
for solutions $u(x,t)$ that satisfy suitable boundary conditions posed at $\partial\Omega$. 
In the case when $\Omega=\Rnum$ is the whole spatial domain, 
then $C$ will be a constant of motion 
under suitable spatial decay conditions on solutions $u(x,t)$ as $|x|\to\infty$. 

A conservation law is {\em locally trivial} if 
\begin{equation}\label{locallytriv}
\dens|_\Esp=D_x\triv,
\quad 
\flux|_\Esp =-D_t\triv
\end{equation} 
hold for all solutions $u(x,t)$,
where $\triv$ is a function depending on $t$, $x$, $u$, and derivatives of $u$,
since then the balance equation for $C$ holds identically
(namely, it contains no useful information about solutions).
Two conservation laws that differ by a trivial conservation law are said to be 
{\em locally equivalent}.
Consequently, only non-trivial conservation laws up to local equivalence are of interest. 

A conserved current $(\dens,\flux)$ can be taken to be a function of only 
$u$, $u_t$, and their $x$-derivatives, in addition to $t$, $x$, 
since up to local equivalence, 
$u_{tt}$ and its derivatives can be eliminated through the given wave equation. 
This will be the situation hereafter. 
As a result, 
every non-trivial conservation law \eqref{conslaw} 
can be shown to arise from a multiplier satisfying the total divergence equation \eqref{multr.eqn}. 
From this equation, 
it is straightforward to see that multipliers and conserved quantities 
are directly related by  \cite{AncBlu2002b}
\begin{equation}\label{Q.relation}
Q=\delta C/\delta u_t = \hat E_{u_t}(\dens)
\end{equation}
where 
$\hat E_{u_t} = \partial_{u_t} -D_x\partial_{u_{tx}} + D_x^2 \partial_{u_{txx}} - \cdots$ 
denotes the spatial Euler operator with respect to $u_t$. 
This relationship implies that $Q$ is a function of only 
$u$, $u_t$, and their $x$-derivatives, in addition to $t$, $x$. 
Therefore, the following main correspondence holds. 

\begin{prop}\label{prop:conslaw.multr}
For any damped wave equation \eqref{waveeqn.gen}, 
there is a one-to-one correspondence between 
non-trivial conservation laws \eqref{conslaw} (up to local equivalence)
and non-zero multipliers \eqref{Q.relation},
where, without loss of generality, $(\dens,\flux)$ and $Q$ are functions of 
$u$, $u_t$, and their $x$-derivatives, in addition to $t$, $x$.  
\end{prop}

\subsection{Variational symmetries} 

An infinitesimal {\em variational symmetry} of a wave equation \eqref{waveeqn.gen} 
is a generator 
\begin{equation}\label{varsymm}
\hat\X= P\partial_u
\end{equation}
where $P$ is a function of $t$, $x$, $u$, and derivatives of $u$ such that 
the action of the prolonged generator preserves the Lagrangian \eqref{Lagr.gen} 
modulo a total space-time divergence. 
Prolongation of a generator \eqref{varsymm} to act on derivatives of $u$ 
is simply given by the corresponding total derivatives of $P$: 
$\pr\hat\X = P\partial_u + D_x P\partial_{u_x} + D_t P\partial_{u_t} +\cdots$.
This formulation of variational symmetries comprises Lie point symmetries, 
contact symmetries, as well as higher order symmetries. 

The variational symmetry condition is explicitly given by 
\begin{equation}\label{varsymm.Lagr.eqn}
\begin{aligned}
\pr\hat\X(L^u)  & = D_t K^t + D_x K^x \\
& = \big( {-}u_t D_t P + c^2 u_x D_x P +g(t,x,u)P \big)e^{2A}
\end{aligned}
\end{equation}
for some functions $K^t$, $K^x$ of $t$, $x$, $u$ and derivatives of $u$. 
Integration by parts applied to the second equality yields 
\begin{equation}\label{varsymm.eqn}
\big(u_{tt} + a(t) u_t + g(t,x,u) -c^2 u_{xx} \big)e^{2A} P
= D_t(K^t - \Gamma^t) + D_x(K^x - \Gamma^x)
\end{equation}
where
\begin{equation}
\Gamma^t =-e^{2A} u_t P ,
\quad
\Gamma^x =e^{2A} u_x P .
\end{equation}
Observe that equation \eqref{varsymm.eqn} coincides with a multiplier equation \eqref{multr.eqn} 
in which the conserved density and spatial flux are given by 
\begin{equation}
\dens= K^t - \Gamma^t,
\quad
\flux= K^x - \Gamma^x
\end{equation}
Thus, the following correspondence is established. 

\begin{prop}\label{prop:multr.varsymm}
For any damped wave equation \eqref{waveeqn.gen}, 
$Q=e^{2A}P$ is a multiplier iff $\hat\X =P\partial_u$ is variational symmetry,
where $e^{2A}$ is a variational integrating factor given by expression \eqref{A}.
\end{prop}

As a consequence, 
a one-to-one correspondence holds between non-trivial variational symmetries
and non-trivial conservation laws \eqref{conslaw} (up to local equivalence). 
This constitutes a statement of Noether's theorem. 

A variational Lie point symmetry has the form 
\begin{equation}\label{varsymm.point}
\hat\X =\big( \eta(t,x,u) - \tau(t,x,u)u_t - \xi(t,x,u)u_x \big)\partial_u, 
\end{equation}
which is well known to be equivalent to a canonical generator \cite{Olv-book,BA-book}
\begin{equation}\label{varsymm.point.canonical}
\X = \eta(t,x,u)\partial_u + \tau(t,x,u)\partial_t + \xi(t,x,u)\partial_x . 
\end{equation}
This corresponds to a linear first-order multiplier 
\begin{equation}
Q= Q_0(t,x,u) + Q_1(t,x,u)u_t + Q_2(t,x,u)u_x
\end{equation}
via 
$Q_0 = e^{2A}\eta$, $Q_1 = -e^{2A}\tau$, $Q_2 = -e^{2A}\xi$. 
Likewise, 
a general first-order multiplier $Q(t,x,u,u_t,u_x)$ corresponds to 
a variational contact symmetry $\hat\X =P(t,x,u,u_t,u_x)\partial_u$
with the canonical form \cite{BA-book}
\begin{equation}\label{varsymm.contact.canonical}
\X = \big( P - u_t P_{u_t} - u_x P_{u_x} \big)\partial_u -P_{u_t}\partial_t -P_{u_x}\partial_x .
\end{equation}

\subsection{Determining equations}

For a given damped wave equation \eqref{waveeqn.gen},
all multipliers $Q$ are determined by the condition \eqref{multr.eqn}. 
This can be formulated efficiently by use of the Euler operator 
$E_{u} = \partial_{u} -D_t\partial_{u_{t}} -D_x\partial_{u_{x}} + D_t^2 \partial_{u_{tt}} + D_tD_x \partial_{u_{tx}} + D_x^2 \partial_{u_{xx}} - \cdots$ 
which has the property that it annihilates a function 
iff the function is a total divergence. 
Consequently, 
the determining condition \eqref{multr.eqn} is equivalent to 
the Euler operator equation
\begin{equation}\label{multr.deteqn}
E_u\big( (u_{tt} + a(t) u_t + g(t,x,u) -c^2 u_{xx})Q\big) =0 . 
\end{equation}
Similarly, since all variational symmetries $\hat\X = P\partial_u$ 
are determined by the condition \eqref{varsymm.Lagr.eqn}, 
an equivalent formulation is provided by the Euler operator equation
\begin{equation}\label{varsymm.deteqn}
E_u\big( \hat\X(L^u) \big)=0 . 
\end{equation}

Note that these respective {\em determining equations} \eqref{multr.deteqn} and \eqref{varsymm.deteqn} 
hold off of the solution space $\Esp$ of the wave equation \eqref{waveeqn.gen}. 
They are equivalent to each other, 
due to the correspondence between multipliers and variational symmetries 
stated in Proposition~\ref{prop:multr.varsymm}. 
This can be seen directly by substituting equation \eqref{varsymm.eqn} 
into equation \eqref{varsymm.Lagr.eqn} and using the correspondence relation 
\begin{equation}\label{Q.P.rel}
Q = e^{2A} P . 
\end{equation}

Any class of multipliers (such as linear first-order),
or class of variational symmetries (such as Lie point type), 
determines a corresponding class of local conservation laws
whose conserved densities have the form $\dens = \int Q\, du_t + F$, 
up to local equivalence, 
for some function $F$ with no dependence on $u_t$ (and its derivatives). 

For a class of multipliers or variational symmetries or conserved densities of interest, 
the respective determining equation for $Q$ or $P$ yields an overdetermined system 
after the equation is split with respect to variables that do not appear respectively 
in $Q$ or $P$. 
The solutions of the resulting system then yield all multipliers or variational symmetries
belonging to the given class. 
For each solution, a corresponding non-trivial conservation law 
arises through equation \eqref{multr.eqn},
where an explicit form for $(\dens,\flux)$ can be found by various methods 
directly in terms of $Q$ or $P$. 
(See \Ref{Anc-review} for a summary). 

As a final remark,
since $\hat\X=P\partial_u$ preserves the extremals of $L^u$,
note that it preserves the solution space $\Esp$ of the damped wave equation \eqref{waveeqn.gen}
and thus is an infinitesimal symmetry. 
This implies that the variational symmetry determining equation \eqref{varsymm.deteqn}
splits into a system comprising the symmetry determining equation
\begin{equation}\label{symm.deteqn}
\big( D_t^2 P + a(t) D_tP + g_u(t,x,u)P -c^2 D_x^2 P \big)\big|_\Esp =0
\end{equation}
plus extra equations that are necessary and sufficient for a symmetry
to be variational. 
Similarly, the multiplier determining equation \eqref{multr.deteqn}
can be shown to split into a system comprised by
the adjoint of symmetry determining equation
\begin{equation}\label{adjsymm.deteqn}
\big( D_t^2 Q -D_t(a(t)Q) + g_u(t,x,u)Q  -c^2 D_x^2 Q \big)\big|_\Esp =0
\end{equation}
plus extra equations.
Solutions of the adjoint equation \eqref{adjsymm.deteqn} are known
as adjoint-symmetries \cite{AncBlu2002a,AncBlu2002b,AncBlu1997},
and hence a multiplier is an adjoint-symmetry that satisfies extra conditions.
In the case of an undamped wave equation \eqref{new.waveeqn}, 
adjoint-symmetries coincide with symmetries,
and multipliers coincide with variational symmetries. 
The preceding formulation has been extensively developed 
in \Ref{Anc-review,Anc2022}.

\section{Wave equations with removable damping}\label{sec:equiv}

An interesting question is whether 
any damped nonlinear wave equations in the class \eqref{waveeqn} 
can be invertibly transformed into an undamped equation in the same class. 
This is a particular case of the general equivalence problem 
for this class of wave equations. 
By inspection, note that the set of general equivalence transformations 
includes scalings $u\to \lambda u$, $\lambda=\const\neq 0$, 
and shifts $u\to u+\epsilon$, $\epsilon=\const$. 

As shown in Section~\ref{sec:prelims}, 
the change of variable \eqref{newvar} transforms 
the class of damped wave equations \eqref{waveeqn.gen} 
into an equivalent class of undamped wave equations \eqref{new.waveeqn} 
with a different self-interaction term \eqref{new.selfinteraction}. 
This change of variable can be generalized by a time-dependent shift 
\begin{equation}\label{newvar.gen}
u= e^{-A}v +f(t)
\end{equation}
with $A$ given by expression \eqref{A}. 
A scaling on $u$ is also permitted, but it can be adsorbed into 
the arbitrary constant appearing in $A$. 
Applying the change of variable \eqref{newvar.gen} 
to the translation-invariant damped nonlinear wave equations \eqref{waveeqn} 
shows that the self-interaction term $g(u)$ is transformed into the more general form
\begin{equation}\label{transform.g}
h(t,v)=  -(\tfrac{1}{2}a_t + \tfrac{1}{4}a^2)v + e^{A}\big( a f_t + f_{tt} + g(e^{-A}v+f) \big)
\end{equation}
in which $t$ can appear explicitly. 
The condition under which $h(t,v)$ will be translation invariant is given by 
$h_t(t,v)\equiv 0$,
which can be written out explicitly in terms of the original self-interaction $g(u)$ as 
\begin{equation}\label{equiv.deteqn}
0 = 
g -(u -f -2f_t/a) g_u  - (a_t +a_{tt}/a)(u -f) + (a + 2a_t/a) f_t + 3 f_{tt} + (2/a) f_{ttt} . 
\end{equation}
This determining condition \eqref{equiv.deteqn} is straightforward to solve 
by separation of variables:
\begin{equation}\label{g.a.ODEs}
f(t)+ 2 f'(t)/a(t) = \mu,
\quad
g'(u) - g(u)/(u-\mu) = \kappa = -(a'(t) + a''(t)/a(t)),
\quad
\mu,\kappa=\const
\end{equation}
which yields a linear first-order ODE on $g$ and a nonlinear second-order ODE on $a$,
along with 
\begin{equation}\label{f}
f= \nu e^{-A}+\mu,
\quad
\nu=\const
\end{equation}
Substitution of $f$ into the change of variable \eqref{newvar.gen} yields
\begin{equation}\label{newvar.shift}
u= e^{-A}(v +\nu) +\mu . 
\end{equation}
This is the most general transformation that removes the damping term $a(t)u_t$. 
Note that both $\mu$ and $\nu$ can be put equal to $0$ 
modulo a shift of $u$ and $v$. 

The ODE \eqref{g.a.ODEs} for $g$, with $\mu=\nu=0$, is easily integrated to get 
\begin{equation}\label{g.sol}
g(u) = (\sigma + \kappa \ln|u|)u,
\quad
\sigma,\kappa=\const .
\end{equation}
Substitution of this expression into the transformed self-interaction term \eqref{transform.g}, 
with $\mu=\nu=0$, 
yields
$h(t,v)= (\sigma -\tfrac{1}{2}a_t -\tfrac{1}{4}a^2 -\kappa A  + \kappa \ln|v|)v$. 
The terms involving $a$ and $A$ combine into a constant 
via the ODE \eqref{g.a.ODEs} for $a$,
which gives
\begin{equation}\label{a.eqn}
\tfrac{1}{2}\kappa\int^{t}_{t_0} a(t)\,dt + \tfrac{1}{4} a(t)^2 + \tfrac{1}{2} a'(t) 
= \sigma_0 =\const
\end{equation}
Hence, the term \eqref{transform.g} can be expressed in the simplified form 
\begin{equation}\label{g.equiv}
h(t, v) = (\sigma -\sigma_0 + \kappa\ln|v|)v
\end{equation}
which has no explicit dependence on $t$ as desired. 

The preceding analysis establishes the following equivalence result. 

\begin{thm}\label{thm:removabledamping}
A damped nonlinear wave equation $u_{tt} + a(t) u_t + g(u) =c^2 u_{xx}$
can be invertibly transformed into an undamped nonlinear wave equation 
$\tilu_{tt} + \tilg(\tilu) =c^2 \tilu_{xx}$
with $\tilu = \exp(\tfrac{1}{2}\int^t_{t_0} a(t)\,dt)u$, modulo shifts, 
iff $g(u)$ has the form \eqref{g.sol} 
and $a(t)$ satisfies equation \eqref{a.eqn}. 
The transformed self-interaction term is given by 
$\tilg(\tilu) = g(\tilu)\big|_{\sigma=\tilde \sigma}$ 
where $\tilde\sigma = \sigma -\sigma_0$ is a shift. 
\end{thm}

This constitutes a non-trivial equivalence transformation within 
the class of wave equations \eqref{waveeqn}. 

An explicit form for $a(t)$ is given by 
\begin{equation}\label{a.sol}
\int^{a(t)}_{a_0} \frac{dy}{W\big(\alpha e^{y^2/(2\kappa)}\big) + 1} 
=\kappa (t_0-t) ,
\quad
a_0, t_0, \alpha =\const
\end{equation}
where $W$ denotes the Lambert function. 
This quadrature is derived as follows. 
The second-order ODE \eqref{g.a.ODEs} for $a$ is invariant under translations in $t$, 
whereby its order can be reduced by expressing $a'(t) = b(a(t))$, 
which leads to a first-order ODE $b'(a)= -(1+\kappa/b(a)) a$. 
Since this ODE is separable, it can be integrated to get 
$b(a) +\kappa - \kappa\ln(b(a)+\kappa) = \alpha - \tfrac{1}{2}a^2$,
where $\alpha$ is an arbitrary constant. 
The solution of this algebraic equation is given by 
$b(a) = -\kappa\big( 1+W( \alpha\exp(\tfrac{1}{2\kappa}y^2) ) \big)$
in terms of the Lambert $W$ function,
after a redefinition of the arbitrary constant. 
Finally, integration of the separable ODE $a'(t) = b(a(t))$ 
then yields the result \eqref{a.sol}.

Note that $\sigma_0$ is determined implicitly in terms of $\alpha$ and $a_0$
through combining equations \eqref{a.sol} and \eqref{a.eqn}. 
In detail, 
the derivative of equation \eqref{a.eqn} yields $\kappa a(t) = -(a(t)a'(t) + a''(t))$. 
Integration from $t_0$ to $t$ then leads to 
$\sigma_0 = \tfrac{1}{4}a(t_0)^2 + \tfrac{1}{2}a'(t_0)$
after equation \eqref{a.eqn} has been substituted. 
Differentiation of equation \eqref{a.sol} with respect to $t$, 
followed by evaluation at $t=t_0$, gives the relation 
$a'(t_0) = -\kappa\big( 1+W(\alpha\exp(a(t_0)^2/(2\kappa))) \big)$, 
where $a(t_0) =a_0$. 
Hence, 
\begin{equation}\label{shift.const}
\sigma_0 = \tfrac{1}{4}a_0{}^2 -\tfrac{1}{2}\kappa\big( 1+W(\alpha\exp(\tfrac{1}{2\kappa}a_0{}^2)) \big).
\end{equation}

In the special case $\alpha=0$, the quadrature \eqref{a.sol} can be evaluated
using $W(0)=0$ to get the explicit expression $a(t)=a_0 +\kappa(t_0-t)$.
Its qualitative features are shared by $a(t)$ in the general case $\alpha\neq0$: 
for $\kappa<0$, $a(t)$ is increasing for all $t$; 
for $\kappa>0$, $a(t)$ is decreasing, reaches $0$ in a finite time, say $t=t_1$,
and can be extended as a piecewise function which is $0$ for all $t>t_1$.

\subsection{Conservation laws} 

The local conservation laws of the wave equation 
\begin{equation}\label{waveeqn.ln.nonlin}
u_{tt} + a(t) u_t +\sigma u + \kappa u\ln|u| =c^2 u_{xx}
\end{equation} 
with the damping $a(t)$ given by expression \eqref{a.sol}
will now be derived from applying Noether's theorem to 
the variational symmetries of the equivalent undamped wave equation
\begin{equation}\label{waveeqn.ln.nonlin.undamped}
\tilu_{tt} + (\sigma -\sigma_0) \tilu + \kappa \tilu\ln|\tilu| =c^2 \tilu_{xx}
\end{equation} 
with the constant $\sigma_0$ given by expression \eqref{shift.const}. 
The equivalence transformation that removes the damping is $u = e^{-A}\tilu$, 
with $A$ given by expression \eqref{A}. 

If the coefficients of both linear terms $\sigma u$ and $(\sigma -\sigma_0)\tilu$ 
are positive, 
namely $\sigma=m^2$ and $\sigma -\sigma_0=\tilde m^2$, 
then these physically represent a mass term.  
In this situation, the wave equation \eqref{waveeqn.ln.nonlin} will describe 
waves with mass $m$ which have time-dependent damping $a(t)$ 
and weak (logarithmic) nonlinear self-interaction $k u\ln|u|$. 
The equivalent wave equation \eqref{waveeqn.ln.nonlin.undamped} 
describes undamped waves with the same self-interaction 
but a different mass. 

It is convenient to recall the conservation laws possessed by 
a general undamped nonlinear wave equation 
\begin{equation}\label{undamped.waveeqn}
\tilu_{tt} + \tilg(\tilu) =c^2 \tilu_{xx}
\end{equation} 
where $\tilg(\tilu)$ is arbitrary. 
The variational symmetries of this wave equation 
\cite{Whitham-book,AncBlu2002a,Strauss-book}
are the span of, in canonical form, 
a $t$-translation $\X = \partial_t$, 
a $x$-translation $\X = \partial_x$, 
and a Lorentz boost $\X = x\partial_t + c^2 t\partial_x$. 
These are point symmetries that generate the (continuous) Poincar\'e group 
$\Rnum^2\rtimes SO(1,1)$ in two dimensional space-time. 
Correspondingly, by Noether's theorem, 
the characteristic form of these three symmetries yield the respective multipliers 
\begin{equation}\label{undamped.multrs}
Q=\tilu_t,
\quad
Q=\tilu_x,
\quad
Q = x \tilu_t + c^2 t \tilu_x . 
\end{equation}
The resulting conservation laws are given by the conserved currents
\begin{align}
& 
\dens = \tfrac{1}{2} (\tilu_t{}^2 +c^2 \tilu_x{}^2) + \tilG(\tilu), 
&&
\flux = -c^2 \tilu_t \tilu_x ;
\label{undamped.conslaw.ener}\\
& 
\dens = \tilu_t \tilu_x ,
&&
\flux= -\tfrac{1}{2} (\tilu_t{}^2 +c^2 \tilu_x{}^2) +\tilG(\tilu) ;
\label{undamped.conslaw.mom}\\
& 
\dens = \tfrac{1}{2}x (\tilu_t{}^2 +c^2 \tilu_x{}^2) + c^2 t \tilu_t \tilu_x + x \tilG(\tilu), 
&&
\flux = -c^2\big( x \tilu_t \tilu_x +\tfrac{1}{2} t (\tilu_t{}^2 +c^2 \tilu_x{}^2) -t \tilG(\tilu)  \big) ;
\label{undamped.conslaw.boostmom}
\end{align}
with $\tilG(\tilu)=\smallint \tilg(\tilu)\, d\tilu$. 
Physically, they describe conservation of energy, momentum, and boost momentum;
$\tilG(\tilu)$ represents a potential energy term. 

For use in Section~\ref{sec:results}, it is also worthwhile to recall that 
the wave equation \eqref{undamped.waveeqn} admits additional 
conservation laws when $\tilg(\tilu)\equiv 0$. 
This is commonly called a massless Klein-Gordon equation in the physics literature. 
Its variational point symmetries, in canonical form, are given by \cite{Ibr-book,PopChe}
the span of $\X_\pm = f_\pm (x \pm ct)(\partial_t \pm c\partial_x)$,
where $f_\pm(x\pm ct)$ is an arbitrary function.  
These symmetries preserve the light cones $(x-x_0)^2 -c^2 (t-t_0)^2 =0$ 
centered at all points $(x_0,t_0)$ in space-time. 
The symmetry subspace such that $f_\pm(x\pm ct)$ is at most quadratic 
comprises the previous generators of the translations and the Lorentz boost
plus $\X = t\partial_t + x\partial_x$, which generates a dilation, 
$\X = (c^2 t^2+x^2)\partial_t + 2c^2 tx\partial_x$ 
and $\X = 2tx\partial_t + (t^2+x^2)\partial_x$, 
which generate conformal (inversion) transformations. 
These six geometric symmetries together generate 
the (continuous) Lorentzian conformal group in two dimensional space-time. 
By Noether's theorem, 
the characteristic form of the symmetries $\X_\pm$ constitute multipliers 
$Q=f_\pm (x \pm ct) (\tilu_t \pm c\tilu_x)$, 
which yield the set of conservation laws
\begin{equation}\label{free.conslaw.nullener}
\dens = \pm \flux = \tfrac{1}{2} f_\pm(x \pm ct)t (\tilu_t \pm c \tilu_x)^2 . 
\end{equation} 
They include, when $f_\pm(x\pm ct)$ is at most quadratic,
the previous conservation laws for 
energy \eqref{undamped.conslaw.ener}, momentum \eqref{undamped.conslaw.mom},
and boost momentum \eqref{undamped.conslaw.boostmom}, 
with $\tilG(\tilu)\equiv 0$,
plus the following three additional conservation laws:
\begin{align}
& 
\dens = \tfrac{1}{2} t (\tilu_t{}^2 +c^2 \tilu_x{}^2) +x\tilu_t \tilu_x , 
&&
\flux = -\tfrac{1}{2} x (\tilu_t{}^2 +c^2 \tilu_x{}^2) -c^2 t \tilu_t \tilu_x ;
\label{free.conslaw.dilener}\\
& 
\dens = \tfrac{1}{2}(x^2 +c^2 t^2) (\tilu_t{}^2 +c^2 \tilu_x{}^2) + 2 c^2 tx \tilu_t \tilu_x , 
&&
\flux = -c^2\big( (x^2+ c^2t^2)  \tilu_t \tilu_x + tx (\tilu_t{}^2 +c^2 \tilu_x{}^2)  \big) ;
\label{free.conslaw.confener}\\
& 
\dens = tx (\tilu_t{}^2 +c^2 \tilu_x{}^2) + (x^2 +c^2 t^2)  \tilu_t \tilu_x , 
&&
\flux = -c^2\big( \tfrac{1}{2}(x^2+ c^2t^2)  (\tilu_t{}^2 +c^2 \tilu_x{}^2)  +2tx  \tilu_t \tilu_x \big) ; 
\label{free.conslaw.confmom}
\end{align}
whose multipliers are respectively given by 
$Q=t\tilu_t+x\tilu_x$, 
$Q=(c^2 t^2+x^2)\tilu_t + 2c^2 tx\tilu_x$ 
and $Q=2c^2 tx\tilu_t + (c^2 t^2+x^2)\tilu_x$. 
These three conservation laws physically describe 
a dilational energy \eqref{free.conslaw.dilener}, 
a conformal energy \eqref{free.conslaw.confener},
and a conformal momentum \eqref{free.conslaw.confmom}. 

Returning to the undamped nonlinear wave equation \eqref{waveeqn.ln.nonlin.undamped}, 
the self-interaction term is $\tilg(\tilu) = (\sigma -\sigma_0 +\kappa\ln|\tilu|)\tilu$. 
Hence, this wave equation possesses 
the energy, momentum, and boost momentum conservation laws \eqref{undamped.conslaw.ener}--\eqref{undamped.conslaw.boostmom}
where the potential energy term is given by 
\begin{equation}
\tilG(\tilu) = \tfrac{1}{2}( \sigma -\sigma_0 -\tfrac{1}{2}\kappa +\kappa \ln|\tilu|) \tilu^2 .
\end{equation}
Under the inverse equivalence transformation $\tilu= e^{A}u$ 
which maps this wave equation into the damped wave equation \eqref{waveeqn.ln.nonlin}, 
these conservation laws yield the following equivalent conservation laws: 
\begin{align}
&
\dens = e^{2A}\big( 
\tfrac{1}{2} ((u_t +\tfrac{1}{2}a u)^2 +c^2 u_x{}^2 +(\kappa A -\sigma_0) u^2) + G(u) 
\big),
\quad
\flux = -c^2 e^{2A} (u_t +\tfrac{1}{2}au) u_x ; 
\label{equiv.conslaw.ener}\\
&
\dens = e^{2A} u_t u_x ,
\quad
\flux= -e^{2A} \big( \tfrac{1}{2}(u_t{}^2 +c^2 u_x{}^2) - G(u) \big) ; 
\label{equiv.conslaw.mom}\\
&\begin{aligned}
\dens & = e^{2A}\big( \tfrac{1}{2}x ((u_t+\tfrac{1}{2}a u)^2 +c^2 u_x{}^2 + (\kappa A -\sigma_0) u^2) + c^2 t u_t u_x + x G(u) \big), 
\\
\flux & = -e^{2A}c^2 \big( 
x (u_t +\tfrac{1}{2} a u) u_x +\tfrac{1}{2} t (u_t{}^2 +c^2 u_x{}^2) -t G(u) 
-\tfrac{1}{4}a u^2 \big) ; 
\end{aligned}
\label{equiv.conslaw.boostmom}
\end{align}
where 
\begin{equation}\label{equiv.potener}
G(u)=\tfrac{1}{2}( \sigma -\tfrac{1}{2}\kappa +\kappa\ln|u| ) u^2
\end{equation}
is the potential energy term. 
All of these conservation laws have been simplified by use of relation \eqref{a.eqn},
and the addition of a locally trivial term \eqref{locallytriv} in the latter two. 

For completeness, it is worthwhile to determine if 
the damped nonlinear wave equation \eqref{waveeqn.ln.nonlin}
possesses any additional conservation laws arising from 
variational point or contact symmetries that are not shared by 
the equivalent wave equation \eqref{undamped.waveeqn} 
with an arbitrary self-interaction. 
Through the equivalence transformation $u= e^{-A}\tilu$
combined with the Noether correspondence in Proposition~\ref{prop:multr.varsymm}, 
variational symmetries $\hat\X = P(t,x,\tilu,\tilu_t,\tilu_x)\partial_{\tilu}$ 
correspond to first-order multipliers $Q=P(t,x,\tilu,\tilu_t,\tilu_x)$,
which are determined by 
\begin{equation}\label{multr.deteqn.ln.nonlin}
E_{\tilu}\big( (\tilu_{tt} + (\sigma -\sigma_0) \tilu + \kappa \tilu\ln|\tilu| -c^2 \tilu_{xx})
P(t,x,\tilu,\tilu_t,\tilu_x) \big) =0
\end{equation}
where $E_{\tilu}$ is the Euler operator.  
This determining equation splits with respect to 
second- and higher- order derivatives of $\tilu$, yielding a system of equations for $P$,
where $\sigma$, $\sigma_0$, $\kappa$ are taken to be arbitrary. 
It is fairly straightforward to solve this system by use of Maple 
(as discussed in the Appendix). 
The result is that the only admitted multipliers $Q=P$ 
are the span of the ones \eqref{undamped.multrs} given by 
the two translations and the Lorentz boost. 
This establishes the following classification. 

\begin{prop}\label{prop:waveeqn.ln.nonlin.conslaws}
For the damped nonlinear wave equation \eqref{waveeqn.ln.nonlin},
all conservation laws arising from variational point and contact symmetries 
are given by the span of the conserved currents \eqref{equiv.conslaw.ener}--\eqref{equiv.potener}. 
\end{prop}

On the spatial domain $\Omega=\Rnum$, 
the resulting conserved quantities describe 
a generalized energy 
\begin{equation}\label{ln.nonlin.ener}
E = \int_\Rnum e^{2A}
\big( 
\tfrac{1}{2} ((u_t +\tfrac{1}{2}a u)^2 +c^2 u_x{}^2 +(\kappa A -\sigma_0) u^2) + G(u) 
\big)\,dx, 
\end{equation}
a generalized momentum 
\begin{equation}\label{ln.nonlin.mom}
M = \int_\Rnum e^{2A} u_t u_x \,dx, 
\end{equation}
and a generalized boost momentum
\begin{equation}\label{ln.nonlin.boostmom}
K = \int_\Rnum e^{2A}\big( 
\tfrac{1}{2}x ((u_t+\tfrac{1}{2}a u)^2 +c^2 u_x{}^2 + (\kappa A -\sigma_0) u^2) + c^2 t u_t u_x + x G(u) 
\big)\,dx . 
\end{equation}
Their physical meaning can be understood by expressing them in a factorized form 
\begin{align}
& E = e^{2A} \tilde E,
&&
\tilde E(t) = \int_\Rnum \big( 
\tfrac{1}{2} ((u_t +\tfrac{1}{2}a u)^2 +c^2 u_x{}^2 +(\kappa A -\sigma_0) u^2) + G(u) 
\big)\,dx, 
\\
& M = e^{2A} \tilde M,
&&
\tilde M(t) = \int_\Rnum u_t u_x \,dx, 
\\
& K = e^{2A} \tilde K,
&&
\tilde K(t) = \int_\Rnum \big( \tfrac{1}{2}x ((u_t+\tfrac{1}{2}a u)^2 +c^2 u_x{}^2 + (\kappa A -\sigma_0) u^2) + c^2 t u_t u_x + x G(u) 
\big)\,dx
\end{align}
where, by comparison with the undamped case, the integrals 
$\tilde E(t)$, $\tilde M(t)$, and $\tilde K(t)$ have the respective forms of 
an energy quantity, a momentum quantity, and a boost momentum quantity. 
Positivity of the damping $a>0$ implies that $e^{2A} = \exp(\int^{t}_{t_0} a\,dt)$ 
is an increasing exponential function of $t$,
and thus the latter quantities are decreasing functions
\begin{equation}
\tilde E(t) = e^{-\int^{t}_{t_0} a\,dt} \tilde E(t_0), 
\quad
\tilde M(t) = e^{-\int^{t}_{t_0} a\,dt} \tilde M(t_0), 
\quad
\tilde K(t) = e^{-\int^{t}_{t_0} a\,dt} \tilde K(t_0), 
\end{equation}
due to $E=E|_{t=t_0}=\tilde E(t_0)$, $M=M|_{t=t_0}=\tilde M(t_0)$, $K=K|_{t=t_0}=\tilde K(t_0)$,
which hold by conservation of the generalized quantities \eqref{ln.nonlin.ener}, \eqref{ln.nonlin.mom}, \eqref{ln.nonlin.boostmom}.

\subsection{Light-cone energies}

A counterpart of the Klein-Gordon conservation laws \eqref{free.conslaw.nullener} 
when the functions $f_\pm$ are taken to be constant 
holds for a general undamped wave equation \eqref{undamped.waveeqn}. 
Specifically, the conservation laws for
energy \eqref{undamped.conslaw.ener} and momentum \eqref{undamped.conslaw.mom} 
can be linearly combined to obtain 
\begin{equation}\label{undamped.conslaw.nullener}
\dens_{\pm} = \tfrac{1}{2} (\tilu_t \pm c \tilu_x)^2 + \tilG(\tilu), 
\quad
\flux_{\pm} = -\tfrac{1}{2} (\tilu_t \pm c \tilu_x)^2 +\tilG(\tilu), 
\end{equation}
modulo a locally trivial conserved current. 
The resulting conserved quantity 
\begin{equation}\label{undamped.nullener}
E_\pm = \int_\Rnum \big( \tfrac{1}{2}(\tilu_t \pm c \tilu_x)^2 + \tilG(\tilu) \big)\, dx 
\end{equation}
describes an energy-momentum associated with the light cone. 
In particular, 
the kinetic term involves $\tilu_t \pm c \tilu_x$ whose form is given by
the derivative $\partial_t \pm c \partial_x$ along the light cone lines $x\pm ct=\const$
\cite{Strauss-book}. 
Note that $\tilu_t\pm c \tilu_x = \delta E_\pm/\delta \tilu_t = \tilde Q_\pm$ 
is also the corresponding multiplier.

This observation is usually omitted in the literature on conservation laws
of semilinear wave equations
(see e.g. \cite{Whitham-book,AncBlu2002a,Strauss-book,BisKarBokZam,RugSpe}).

Since these conserved quantities \eqref{undamped.nullener} hold in particular 
for the undamped nonlinear wave equation \eqref{waveeqn.ln.nonlin.undamped}, 
they yield equivalent conserved quantities 
\begin{equation}\label{ln.nonlin.nullener}
E_\pm = \int_\Rnum e^{2A}\big( 
\tfrac{1}{2} ((u_t +\tfrac{1}{2}a u \pm c u_x)^2 +(\kappa A -\sigma_0) u^2) + G(u) 
\big)\,dx, 
\end{equation}
for the equivalent damped wave equation \eqref{waveeqn.ln.nonlin}. 
The corresponding conserved currents are given by 
\begin{equation}\label{equiv.conslaw.nullener}
\begin{aligned}
\dens_{\pm} & = e^{2A}\big( 
\tfrac{1}{2} ((u_t +\tfrac{1}{2}a u \pm c u_x)^2 +(\kappa A -\sigma_0) u^2) + G(u) 
\big),
\\
\flux_{\pm} & = -c e^{2A} \big( \tfrac{1}{2} ((u_t +\tfrac{1}{2}a u \pm c u_x)^2 -(\kappa A -\sigma_0) u^2) - G(u) \big) .
\end{aligned}
\end{equation}

\section{Conservation laws for intrinsic damping}\label{sec:results}

Hereafter, only nonlinear wave equations \eqref{waveeqn} with intrinsic damping 
that cannot be removed by a transformation of $u$ will be considered. 
From Theorem~\ref{thm:removabledamping}, 
the necessary and sufficient condition is given by 
the negation of equation \eqref{equiv.deteqn},
where $f$ is expression \eqref{f}. 
This requires that $a(t)$ and $g(u)$ satisfy the inequality 
\begin{equation}\label{cond:intrinsic}
g_u - g/(u-\mu) + (a_t + a_{tt}/a) \not\equiv 0
\end{equation} 
for all $\mu=\const$, 
whereby they cannot simultaneously have 
the respective forms \eqref{a.sol} and \eqref{g.sol} modulo a shift on $u$. 

The conserved quantities that typically are admitted by nonlinear wave equations 
are of energy-momentum type whose conserved density is quadratic in $u_t$, $u_x$. 
For example, 
see the densities \eqref{undamped.conslaw.ener}, \eqref{undamped.conslaw.mom}, \eqref{undamped.conslaw.boostmom}
for energy, momentum, and boost momentum of 
the general undamped nonlinear wave equation \eqref{undamped.waveeqn},
as well as their generalizations \eqref{equiv.conslaw.ener}, \eqref{equiv.conslaw.mom}, \eqref{equiv.conslaw.boostmom}
which hold for the particular damped nonlinear wave equation \eqref{waveeqn.ln.nonlin}. 

For a general damped nonlinear equation \eqref{waveeqn},
it is therefore natural to seek all conserved quantities of the form 
\begin{equation}\label{quadratic.integral}
C = \int_\Rnum\big( F_1 u_t{}^2 + F_2 u_x u_t + F_3 u_x{}^2  + F_4 u_t + F_5 u_x + F_0 \big)\,dx
\end{equation}
where each $F_i$, ($i=0,\ldots,5$) is a function of $t$, $x$, $u$. 
The general relationship \eqref{Q.relation} between conserved densities and multipliers
shows that such quantities \eqref{quadratic.integral} 
are characterized by multipliers that have the linear first-order form 
\begin{equation}\label{lin.multr}
Q=2 F_1 u_t + F_2 u_x + F_4 .
\end{equation}
In turn, 
such multipliers correspond to variational point symmetries \eqref{varsymm.point}
whose equivalent canonical form \eqref{varsymm.point.canonical} is given by 
$\eta = e^{-A}F_4$, $\tau = -\tfrac{1}{2}e^{-A}F_1$, $\xi = -e^{A} F_2$, 
due to the Noether correspondence stated in Proposition~\ref{prop:multr.varsymm}. 

The determining equation \eqref{multr.deteqn} for multipliers \eqref{lin.multr}
is given by 
\begin{equation}
E_u\big( (u_{tt} +a(t) u_t -c^2 u_{xx} + g(u))(2F_1 u_t + F_2 u_x + F_4) \big) =0
\end{equation}
where $E_u$ is the Euler operator.  
This equation is required to hold as an identity 
(namely, off of the solution space $\Esp$ of the wave equation \eqref{waveeqn}).
Consequently, it splits with respect to derivatives of $u$, 
and thereby yields a system of equations in which the unknowns consist of
$g(u)$, $a(t)$, $F_1(t,x,u)$, $F_2(t,x,u)$, $F_4(t,x,u)$. 
For the present purpose, 
$g(u)$ must be nonlinear and $a(t)$ must be non-zero:
\begin{equation}\label{cond:nonlin.damped}
g_{uu}\not\equiv 0,
\quad
a\not\equiv 0
\end{equation}
In addition, the intrinsic damping condition \eqref{cond:intrinsic} is imposed. 

The resulting multiplier determining system is nonlinear in the unknowns. 
Its solution requires a lengthy computation, involving several case splittings, 
which can be done interactively with the use of Maple 
(as discussed in the Appendix). 
The following classification result is obtained. 

\begin{prop}\label{prop:lin.multr}
All linear first-order multipliers \eqref{lin.multr} 
admitted by a damped nonlinear wave equation \eqref{waveeqn}
with arbitrary $a(t)$ and $g(u)$ 
are given by 
\begin{equation}\label{case.gen}
Q_{1} = e^{A} u_x .
\end{equation}
Additional linear first-order multipliers are admitted by 
an intrinsically damped nonlinear wave equation \eqref{waveeqn}
only for the following $a(t)$ and $g(u)$ modulo shifts on $u$:
\\
{\rm (i)}
\begin{equation}\label{case1}
a'(t)=0, 
\quad
g(u) = (\tfrac{1}{2}a)^2 u + k u^{1+p},
\quad
k,p=\const
\end{equation}
which admit 
\begin{align}
Q_{2\pm} = & 
e^{a((p+4)c t \pm p x)/(4c)} ( u_t \pm c u_x + \tfrac{1}{2}a u ) ;
\end{align}
{\rm (ii)}
\begin{equation}\label{case2}
a'(t) = -q a(t)^2,
\quad
g(u) = k u^{1 + 4 q},
\quad
k,q=\const
\end{equation}
which admit 
\begin{align}
Q_{3{\rm a}} = & 
\frac{1}{a(t)^{1+1/q}} q u_t  + \frac{1}{a(t)^{1/q}}\big( q x u_x + \tfrac{1}{2} u ) \big), 
\\
Q_{3{\rm b}} = & 
\frac{1}{a(t)^{1+1/q}} q x u_t  +\frac{1}{a(t)^{1/q}} \big( \tfrac{1}{2} xu + \tfrac{1}{2}(q^2x^2 +c^2/a(t)^2) u_x \big) ;
\end{align}
{\rm (iii)}
\begin{equation}\label{case3}
a'(t) = q(a_1{}^2 -a(t)^2),
\quad
g(u) = (\tfrac{1}{2}a_1)^2 u + k u ^{1 + 4q},
\quad
a_1,k,q=\const
\end{equation}
which admit 
\begin{align}
Q_{4\pm} = & 
\frac{e^{\pm a_1q x/c}}{\sqrt{a_1{}^2 - a(t)^2}{\strut}^{1+1/q}} \big( a_1 u_t + a(t) (\pm c u_x + \tfrac{1}{2} a_1 u) \big) . 
\end{align}
\end{prop}

In the latter three cases, the explicit expression for the damping 
$a(t)$, with $a(t_0) = a_0$, is given by:
\\
(i) 
\begin{equation}\label{a.case1}
a(t)=a_0,
\quad
a_0>0 ;
\end{equation}
\\
(ii) 
\begin{equation}\label{a.case2}
a(t) = a_0/(1 +a_0 q (t - t_0)),
\quad
a_0>0 ;
\end{equation}
\\
(iii) 
\begin{equation}\label{a.case3}
\begin{aligned}
a(t) & = a_1\tanh\big(a_1q(t-t_0) +\arctanh(a_0/a_1)\big) \\
& = a_0\big( 1+\tfrac{a_1}{a_0}\tanh(a_1q(t - t_0)) \big)/\big( 1+\tfrac{a_0}{a_1}\tanh(a_1q(t - t_0)) \big),
\quad
a_1>a_0>0 . 
\end{aligned}
\end{equation}

The resulting conservation laws and conserved quantities can be obtained 
straightforwardly by integration of the multiplier equation 
\begin{equation}
(u_{tt} +a(t) u_t  -c^2 u_{xx} + g(u)) Q
= D_x\dens + D_t \flux
\end{equation}
using a descent method in terms of the derivatives of $u$. 
(See \Ref{Anc-review} for details.)

\begin{thm}\label{thm:conslaws}
For intrinsically damped nonlinear wave equations \eqref{waveeqn},
the admitted energy-momentum type conservation laws 
consist of:
\begin{equation}\label{conslaw.gencase}
\dens_{1} = e^{\int^t_{t_0} a(t)\,dt} u_t u_x,
\quad
\flux_{1} = -e^{\int^t_{t_0} a(t)\,dt}\big( 
\tfrac{1}{2}(u_t{}^2  +c^2 u_x{}^2) - \smallint g(u)\,du 
\big) , 
\end{equation}
with arbitrary $a(t)$ and $g(u)$; 
\begin{subequations}\label{conslaw.case1}
\begin{align}
&\begin{aligned}
\dens_{2\pm} & =e^{a_0((p+4)ct \pm p x)/(4c)} \big( 
\tfrac{1}{2}(u_t  \pm cu_x + \tfrac{1}{2}a_0 u)^2 + \tfrac{k}{p+2} u^{p+2} 
\big) , 
\end{aligned}
\\
&\begin{aligned}
\flux_{2\pm} & = \mp ce^{a_0((p+4)ct \pm p x)/(4c)} \big( 
\tfrac{1}{2}(u_t  \pm cu_x + \tfrac{1}{2}a_0 u)^2 - \tfrac{k}{p+2} u^{p+2} 
\big) , 
\end{aligned}                                                                                                                   \end{align}
\end{subequations}
with $a(t)$ and $g(u)$ given by expressions \eqref{case1}, \eqref{a.case1};
\begin{subequations}\label{conslaw.case2a}
\begin{align}
&\begin{aligned}
\dens_{3{\rm a}} & = f_{3}(t)\big( 
\tfrac{1}{2} a(t)^{-1}( u_t^2 + c^2 u_x^2  + \tfrac{k}{2q+1}u^{4q+2} ) 
+ (q x u_x  + \tfrac{1}{2} u) u_t 
\big) , 
\end{aligned}
\\
&\begin{aligned}
\flux_{3{\rm a}} & = -f_{3}(t)\big( 
\tfrac{1}{2}q x ( u_t^2 + c^2 u_x^2  - \tfrac{k}{2q+1} u^{4q+2})  
+ c^2 (a(t)^{-1} u_t  + \tfrac{1}{2} u)u_x 
\big) , 
\end{aligned}
\end{align}
\end{subequations}
and
\begin{subequations}\label{conslaw.case2b}
\begin{align}
&\begin{aligned}
\dens_{3{\rm b}} & = f_{3}(t)\big( 
\tfrac{1}{2} q x a(t)^{-1}(u_t^2 + c^2 u_x^2 + \tfrac{k}{2q+1} u^{4q+2}) 
+ ( \tfrac{1}{2} (q^2 x^2 + c^2 a(t)^{-2} ) u_x  + \tfrac{1}{2} q x u) u_t  
\big) , 
\end{aligned}
\\
&\begin{aligned}
\flux_{3{\rm b}} & = -f_{3}(t)\big( 
\tfrac{1}{4}(q^2 x^2 +c^2 a(t)^{-2})( u_t^2 + c^2 u_x^2 - \tfrac{k}{2q+1} u^{4q+2} )
+ c^2 q x (a(t)^{-1} u_t +\tfrac{1}{2} u) u_x -c^2(\tfrac{1}{2} u)^2 
\big) , 
\end{aligned}
\end{align}
where
\begin{equation}\label{f3}
f_{3}(t) = e^{2A(t)} =1/(q a(t))^{1/q} = (t-t_0+1/(a_0 q))^{1/q}  
\end{equation}
\end{subequations}
with $a(t)$ and $g(u)$ given by expressions \eqref{case2}, \eqref{a.case2};
\begin{subequations}\label{conslaw.case3}
\begin{align}
&\begin{aligned}
\dens_{4\pm} & = e^{\pm a_1q x/c} f_{4}(t)\big(
a_1 ( u_t^2 + c^2 u_x^2 + \tfrac{1}{4}a_1^2(1 -2q)u^2 + \tfrac{k}{2q+1}u^{4q+ 2} )
 + a(t) (a_1 u \pm 2c u_x)u_t 
\big) ,
\end{aligned}
\\
&\begin{aligned}
\flux_{4\pm} & = e^{\pm a_1q x/c} f_{4}(t)\big(
{\mp}c a(t) ( u_t^2 + c^2u_x^2 \pm  c a_1 u u_x  - \tfrac{1}{4} a_1^2(1 + 2q)u^2 
- \tfrac{k}{2q+1} u^{4q + 2} )
-2 a_1 c^2 u_t u_x  
\big) , 
\end{aligned}
\end{align}
where
\begin{equation}\label{f4}
\begin{aligned}
f_{4}(t) = e^{2A(t)}/\sqrt{1- a(t)^2/a_1{}^2}
& = 1/\sqrt{1- a(t)^2/a_1{}^2}{\strut}^{1+1/q} \\
& = \cosh\big(a_1q(t-t_0) +\arctanh(a_0/a_1)\big)^{1+1/q} 
\end{aligned}
\end{equation}
\end{subequations}
with $a(t)$ and $g(u)$ given by expressions \eqref{case3}, \eqref{a.case3}.
\end{thm}

The first conservation law \eqref{conslaw.gencase} has been previously found 
in \Ref{MarBru} for a smaller class of nonlinear wave equations.
The remaining four conservation laws 
\eqref{conslaw.case1}, \eqref{conslaw.case2a}, \eqref{conslaw.case2b}, \eqref{conslaw.case3} 
are new. 

For completeness, 
existence of conservation laws with nonlinear first-order multipliers 
$Q(t,x,u,u_t,u_x)$ 
will be addressed. 
Nonlinearity of $Q$ in the variables $u_t$, $u_x$ implies that 
the corresponding form for conserved quantities will be given by 
a conserved density that is non-quadratic in $(u_t,u_x)$.
The multiplier determining equation \eqref{multr.deteqn} is given by 
\begin{equation}\label{nonlin.multr.deteqn}
E_u\big( (u_{tt} +a(t) u_t -c^2 u_{xx} + g(u))Q(t,x,u,u_t,u_x) \big) =0
\end{equation}
which splits with respect to second- and higher- order derivatives of $u$, 
yielding a system of equations for $Q$, $g$, $a$, 
subject to the same conditions \eqref{cond:intrinsic} and \eqref{cond:nonlin.damped} 
considered for linear multipliers. 
This determining system can be solved straightforwardly by use of Maple 
(as outlined in the Appendix)
and gives the following classification result.

\begin{prop}\label{prop:nonlin.multr}
No nonlinear first-order multipliers are admitted by 
intrinsically damped nonlinear wave equations \eqref{waveeqn}. 
Thus, the only conserved quantities that exist are of 
energy-momentum type \eqref{quadratic.integral}. 
\end{prop}

This further implies that intrinsically damped nonlinear wave equations \eqref{waveeqn.gen} 
do not possess any variational contact symmetries \eqref{varsymm.contact.canonical}
according to Noether correspondence stated in Proposition~\ref{prop:multr.varsymm}.

\subsection{Conserved quantities}

The five energy-momentum type conservation laws \eqref{conslaw.gencase}--\eqref{conslaw.case3}
yield the following conserved quantities:
\begin{equation}
C_{1} = \int_\Rnum e^{\int^t_{t_0} a(t)\,dt} u_t u_x \,dx 
\label{C1}
\end{equation}
for $a(t)$ and $g(u)$ arbitrary; 
\begin{equation}
C_{2\pm} = \int_\Rnum e^{a_0((p+4)ct \pm p x)/(4c)} \big( 
\tfrac{1}{2}(u_t  \pm cu_x + \tfrac{1}{2}a_0 u)^2 + \tfrac{k}{p+2} u^{p+2} 
\big)\, dx 
\label{C2}
\end{equation}
for $a(t)$ and $g(u)$ of the form \eqref{a.case1} and \eqref{case1}; 
\begin{equation}
C_{3{\rm a}} = \int_\Rnum f_{3}(t)\big( 
\tfrac{1}{2} a(t)^{-1}( u_t^2 + c^2 u_x^2  + \tfrac{k}{2q+1}u^{4q+2} ) 
+ (q x u_x  + \tfrac{1}{2} u) u_t 
\big)\, dx ,
\label{C3a}
\end{equation}
and 
\begin{equation}
C_{3{\rm b}} = \int_\Rnum f_{3}(t)\big( 
\tfrac{1}{2} q x a(t)^{-1}(u_t^2 + c^2 u_x^2 + \tfrac{k}{2q+1} u^{4q+2}) 
+ ( \tfrac{1}{2} (q^2 x^2 + c^2 a(t)^{-2} ) u_x  + \tfrac{1}{2} q x u) u_t  
\big)\, dx 
\label{C3b}
\end{equation}
for $a(t)$ and $g(u)$ of the form \eqref{a.case2} and \eqref{case2},
where $f_{3}(t)$ is expression \eqref{f3};
\begin{equation}
C_{4\pm} = \int_\Rnum e^{\pm a_1q x/c} f_{4}(t)\big( 
a_1 ( u_t^2 + c^2 u_x^2 + \tfrac{1}{4}a_1^2(1 -2q)u^2 + \tfrac{k}{2q+1}u^{4q+ 2} )
 + a(t) (a_1 u \pm 2c u_x)u_t 
\big)\, dx 
\label{C4}
\end{equation}
for $a(t)$ and $g(u)$ of the form \eqref{a.case3} and \eqref{case3},
where $f_{4}(t)$ is expression \eqref{f4}.

To understand the physical meaning of these conserved quantities, 
it is useful to consider their equivalent form given by the change of variable \eqref{newvar} 
that invertibly transforms a damped nonlinear wave equation \eqref{waveeqn} 
into an equivalent undamped wave equation \eqref{new.waveeqn}--\eqref{new.selfinteraction}.
The resulting expression for each conserved quantity then can be compared directly
with the well known conserved quantities 
admitted by wave equations 
\begin{equation}\label{undamped.waveeqn.v}
v_{tt} + \tilg(v) = c^2 v_{xx} . 
\end{equation}
Specifically, such wave equations share the same kinetic terms 
as in the undamped wave equations \eqref{new.waveeqn} 
which have a $t$-dependent nonlinearity \eqref{new.selfinteraction}, 
and they also arise from damped wave equations \eqref{waveeqn} 
in the limit where the damping $a(t)$ is taken to vanish such that $u\to v$. 

This comparison will now be carried for each of the five conserved quantities \eqref{C1}--\eqref{C4}.
The explicit form of the change of variable \eqref{newvar} is readily obtained 
by evaluating expression \eqref{A} in terms of the damping \eqref{a.case1}--\eqref{a.case3}. 
For simplicity, it is convenient to take $t_0=0$. 

For conserved quantity \eqref{C1}:
$v= e^{\frac{1}{2}\int^t_{0} a(t)\,dt}u$ yields
\begin{equation}
C_{1} = \int_{\Rnum} v_t v_x\,dx ,
\end{equation}
which is also obtained when $a(t)$ is put equal to $0$ with $u=v$,
where the transformed wave equation is $v_{tt} + g(v) = c^2 v_{xx}$. 
The density in this conserved quantity has the same form 
as the conserved momentum known for wave equation \eqref{undamped.waveeqn.v}, 
namely expression \eqref{undamped.conslaw.mom} with $v=\tilu$. 
Thus, $C_{1}$ describes a generalized momentum. 

For conserved quantity \eqref{C2}:
$v= e^{at/2}u$ yields 
\begin{equation}
C_{2\pm} = \int_\Rnum 
e^{\pm apx/(4c)}\big( e^{apt/4}\tfrac{1}{2} (v_t \pm cv_x)^2 
+ e^{-apt/4} \tfrac{k}{p+2}  v^{p+2} \big)\,dx
\end{equation}
where the transformed wave equation is 
$v_{tt} + k e^{-a pt/2} v^{1+p} = c^2 v_{xx}$. 
Firstly, when $a=0$ and $u=v$, the density in $C_{2\pm}$ reduces to 
$\tfrac{1}{2} (v_t \pm cv_x)^2 + \tfrac{k}{p+2}  v^{p+2}$
which has the same form as the conserved light-cone energies 
known for wave equation \eqref{undamped.waveeqn.v}, 
namely expression \eqref{undamped.conslaw.nullener} with $v=\tilu$. 
Secondly, the exponential factor $e^{\pm a px/(4c)}e^{a pt/4}$
in the kinetic terms is a function of the light-cone variable $x\pm ct$,
which appears in the Klein-Gordon conservation laws \eqref{free.conslaw.nullener}. 
Hence, $C_{2\pm}$ describes generalized light-cone energies. 

For conserved quantities \eqref{C3a} and \eqref{C3b}:
$v=(t+t_1)^{1/(2q)}u$, with $t_1=\tfrac{1}{a_0 q}$, 
yields 
\begin{align}
C_{3{\rm a}} & = \int_\Rnum\big(  
\tfrac{1}{2} (t+t_1)(v_t^2 + v_x^2c^2)+ x v_t v_x  
+ (t+t_1)^{-1} (\tfrac{2q - 1}{8q^2} v^2 + \tfrac{k}{4q+2} v^{4q+2}) \big)\,dx , 
\\
C_{3{\rm b}} & =  \int_\Rnum\big(  
\tfrac{1}{2}x (t+t_1) (v_t^2 + c^2v_x^2)  + \tfrac{1}{2}(x^2 + c^2 (t+t_1)^2) v_t v_x
+ x(t+t_1)^{-1} (\tfrac{2q - 1}{8q^2} v^2 + \tfrac{k}{4q+2} v^{4q+2} \big)\,dx
\end{align}
where 
$v_{tt} + (t+t_1)^{-2}( \tfrac{2q-1}{4q^2} v +  t_1{}^2k v^{4q+1} )= c^2 v_{xx}$
is the transformed wave equation. 
(Note that a constant scaling of $v$ has been made here for simplicity.)
Firstly, 
the limit $a(t)\to0$ is given by $a_0\to 0$ in expression \eqref{a.case2}, 
which implies $t_1\to\infty$. 
This limit is singular in $C_{3{\rm a}}$ and $C_{3{\rm b}}$. 
After a suitable scaling, the densities have the respective limits
$\tfrac{1}{2} (v_t^2 + v_x^2c^2)$ and $\tfrac{1}{2} c^2 v_t v_x$,
which match the conserved energy and momentum densities 
known for wave equation \eqref{undamped.waveeqn.v}, 
namely expressions \eqref{undamped.conslaw.ener} and \eqref{undamped.conslaw.mom} 
with $v=\tilu$. 
However, when $a(t)\not\equiv0$, 
the form of the kinetic terms in the densities do not match any of 
the conserved densities of wave equation \eqref{undamped.waveeqn.v}, 
namely expressions \eqref{undamped.conslaw.ener}, \eqref{undamped.conslaw.mom},
\eqref{undamped.conslaw.boostmom}, \eqref{undamped.conslaw.nullener} 
with $v=\tilu$. 
In particular, the kinetic terms in $C_{3{\rm a}}$ involve factors of $t+t_1$ and $x$,
while the kinetic terms in $C_{3{\rm b}}$ are quadratic in these factors. 
Remarkably, these two patterns appear in the conserved densities for 
dilation energy \eqref{free.conslaw.dilener} and conformal momentum \eqref{free.conslaw.confmom} 
known for the wave equation \eqref{undamped.waveeqn.v} 
in the case $\tilg(v)\equiv 0$. 
Consequently, the quantities $C_{3{\rm a}}$ and $C_{3{\rm b}}$ 
can be viewed as describing a generalized dilational energy 
and a generalized conformal momentum, respectively. 

For conserved quantity \eqref{C4}:
$v=\cosh(a_1q(t+t_1))^{1/(2q)}u$, 
with $t_1=\tfrac{1}{a_1q}\arctanh(\tfrac{a_0}{a_1})$,
yields
\begin{equation}
\begin{aligned}
C_{4\pm} = \int_\Rnum & 
e^{\pm a_1 q x/c} \big( 
\cosh(a_1q(t+t_1)) (v_t^2 + v_x^2c^2)
\pm 2c \sinh(a_1q(t+t_1)) v_tv_x 
\\&\quad
+ \cosh(a_1q(t+t_1))^{-1}( \tfrac{a_1{}^2(1-2q)}{4} v^2 +\tfrac{k}{2q+1} v^{4q+2} )
\big)\,dx
\end{aligned}
\end{equation}
where 
$v_{tt} + \cosh(a_1 q(t+t_1))^{-2}( \tfrac{a_1{}^2(1-2q)}{4} v +  k v^{4q+1} )= c^2 v_{xx}$
is the transformed wave equation. 
(Note that a constant scaling of $v$ has been made here for simplicity.)
Firstly, 
the limit $a(t)\to0$ is given by successively putting $a_0 =0$ and $a_1=0$ 
in expression \eqref{a.case3}, whereby $t_1=0$. 
In this limit, 
the density in $C_{4\pm}$ reduces to 
$\tfrac{1}{2} (v_t^2 + v_x^2c^2) +\tfrac{k}{2q+1} v^{4q+2}$
which matches the form of the conserved energy 
known for wave equation \eqref{undamped.waveeqn.v}, 
namely expression \eqref{undamped.conslaw.ener} with $v=\tilu$. 
But when $a(t)\not\equiv0$, 
the form of the kinetic terms in the densities do not match any of 
the conserved densities of wave equation \eqref{undamped.waveeqn.v}.  
However, the $\cosh$ and $\sinh$ factors in the densities can be expanded out 
in terms of exponentials to get 
\begin{equation}
\begin{aligned}
C_{4\pm} = \int_\Rnum \Big( & 
\sqrt{\tfrac{a_1+a_0}{a_1-a_0}} e^{a_1 q (t \pm x/c)} \big( \tfrac{1}{2} (v_t \pm c v_x)^2 
+ (1-(\tfrac{a(t)}{a_1})^2)( \tfrac{a_1^2(1-2q)}{8} v^2 +\tfrac{k}{4q+2} v^{4q+2} ) \big)
\\&
+\sqrt{\tfrac{a_1-a_0}{a_1+a_0}} e^{-a_1 q (t \mp x/c)} \big( \tfrac{1}{2} (v_t \mp c v_x)^2 
+ (1-(\tfrac{a(t)}{a_1})^2)( \tfrac{a_1^2(1-2q)}{8} v^2 +\tfrac{k}{4q+2} v^{4q+2} )
\big) \Big)\,dx . 
\end{aligned}
\end{equation}
Then the density is a sum of exponentials of the outgoing light-cone variable $x - ct$ 
and the ingoing light-cone variable $x +ct$, respectively multiplied by 
the light-cone energy densities \eqref{undamped.conslaw.nullener} 
known for the wave equation \eqref{undamped.waveeqn.v} with $v=\tilu$. 
Consequently, the quantity $C_{4\pm}$ can be viewed as describing 
a sum of ingoing/outgoing generalized light-cone energies. 
 
The preceding discussion will be reinforced in Section~\ref{sec:symms}
by examining the variational symmetries corresponding to the conserved quantities
\eqref{C1}--\eqref{C4}. 

A final remark is that the conserved quantity \eqref{C2} is singled out by 
the property that it can be expressed in a temporally factorized form
\begin{equation}
C_{2\pm} = e^{a_0(p+4)t/4} \tilde C_{2\pm}(t),
\quad
\tilde C_{2\pm}(t) = \int_\Rnum e^{\pm a_0px/(4c)} \big( 
\tfrac{1}{2}(u_t  + cu_x + \tfrac{1}{2}a_0 u)^2 + \tfrac{k}{p+2} u^{p+2} 
\big)\, dx 
\label{C2.factored}
\end{equation}
where the density of the integral $\tilde C_{2\pm}(t)$ does not explicitly contain $t$. 
Conservation of $C_{2\pm}$ thereby implies that $\tilde C_{2\pm}(t)$ exhibits 
the decay behaviour 
\begin{equation}
\tilde C_{2\pm}(t) = e^{-a_0(p+4)t/4} C_{2\pm}(0) . 
\end{equation}
Note that the other conserved quantities, 
all of which involve a time-dependent damping $a(t)$, 
do not have this property.

\section{Variational symmetries for intrinsic damping}\label{sec:symms}

For the class of damped nonlinear wave equations \eqref{waveeqn.gen}, 
the Noether correspondence given by Propositions~\ref{prop:conslaw.multr} and~\ref{prop:multr.varsymm}
shows that there is a one-to-one correspondence between 
non-trivial conserved quantities \eqref{cons.integral}
and variational symmetries \eqref{varsymm}.
This correspondence is explicitly given by the relation 
\begin{equation}\label{P.C.rel}
P = e^{-2A} \delta C/\delta u_t . 
\end{equation} 

Since an $x$-translation, $\X = \partial_x$ in canonical form, 
is an obvious variational symmetry of the Lagrangian \eqref{Lagr.gen}, 
the main interest is in exploring all of the additional symmetries
and highlighting properties of those that are new or unexpected. 

To proceed, 
the relation \eqref{P.C.rel} will be used to derive 
the variational symmetries corresponding to the conserved quantities
obtained in Proposition~\ref{prop:waveeqn.ln.nonlin.conslaws}
for the nonlinear wave equation \eqref{waveeqn.ln.nonlin} with removable damping,
and in Theorem~\ref{thm:conslaws}
for nonlinear wave equations \eqref{waveeqn.gen} with intrinsic damping. 
For each symmetry, the one-parameter transformation group which it generates, 
and its physical meaning are presented. 
The group parameter will be denoted $\epsilon$,
with $\epsilon=0$ corresponding to the identity transformation. 

The generalized momentum quantity \eqref{C1} 
is conserved for all wave equations \eqref{waveeqn.gen}, 
regardless of the form of the damping $a(t)$ and nonlinearity $g(u)$. 
It has $\delta C_{1}/\delta u_t = e^{\int^t_{t_0} a(t)\,dt} u_x$,
which yields $P _{1} = u_x$. 
Hence, the variational symmetry in characteristic form is $\hat\X_1 = u_x$
which is an $x$-translation with the canonical form 
$\X = -\partial_x$.
This generates the symmetry transformation group $x\to x-\epsilon$. 

From Theorem~\ref{thm:conslaws}, 
there are four remaining conserved quantities to consider. 
Hereafter, $t_0=0$ for simplicity. 

The first quantity is the pair of generalized light-cone energies \eqref{C2} 
which hold for $a(t)$ and $g(u)$ given by expressions \eqref{case1}, \eqref{a.case1}.
For these two energies, 
\begin{equation} 
\delta C_{2\pm}/\delta u_t = e^{a_0((p+4)ct \pm px)/(4c)} (u_t  \pm cu_x + \tfrac{1}{2}a_0 u)
\end{equation}
gives
\begin{equation}\label{P.case1}
P_{2\pm} = e^{a_0p(c t \pm x)/(4c)} ( u_t \pm c u_x + \tfrac{1}{2}a_0 u )
\end{equation}
since $e^{2A} = e^{a t}$. 
This yields the pair of variational symmetries
\begin{equation}
\X_{2\pm} = -e^{a_0p(c t \pm x)/(4c)}\big( 
\partial_t \pm c\partial_x - \tfrac{1}{2}a_0 u\partial_u \big) 
\end{equation}
in canonical form. 
It is useful here to define the light-cone variables
\begin{equation}\label{lightcone.vars}
\zeta_\pm = ct \pm x . 
\end{equation}
Then the variational symmetries take the simpler form 
\begin{equation}\label{symm.case1}
\X_{2\pm} = -e^{a_0 p\zeta_ \pm/(4c)}\big( 
\partial_{\zeta_\pm} - \tfrac{a_0}{4c} u\partial_u \big) , 
\end{equation}
which describes a translation on an exponential of the light-cone variables,
combined with a non-rigid scaling on $u$. 
Explicitly, 
the symmetry transformation groups generated by this pair of symmetries 
can expressed as 
\begin{equation}\label{symmgroup.case1}
e^{-a_0p\zeta_ \pm/(4c)} \to e^{-a_0p\zeta_ \pm/(4c)} + \epsilon,
\quad
u \to (1+\epsilon e^{a_0p\zeta_ \pm/(4c)})^{1/p} u,
\end{equation}
while $\zeta_\mp$ and $e^{a_0\zeta_ \pm/(4c)}u$ are invariant. 
Here the exponential $e^{-a_0 p\zeta_ \pm/(4c)}$ has the role of a canonical coordinate, 
which is a feature not seen typically in variational symmetries 
related to the light-cone for nonlinear wave equations 
(see e.g. \Ref{Whitham-book,AncBlu2002a,Strauss-book,BisKarBokZam,RugSpe}).

The next two conserved quantities are the generalized dilational energy \eqref{C3a} 
and the generalized conformal momentum \eqref{C3b},
both of which hold for $a(t)$ and $g(u)$ given by expressions \eqref{case2}, \eqref{a.case2}.
These two quantities respectively have
\begin{align}
\delta C_{3{\rm a}}/\delta u_t & = 
f_{3}(t)\big( q a(t)^{-1} u_t + q x u_x  + \tfrac{1}{2} u \big) , 
\\
\delta C_{3{\rm b}}/\delta u_t & = 
f_{3}(t)\big( q x a(t)^{-1} u_t + \tfrac{1}{2} (q^2 x^2 + c^2 a(t)^{-2} ) u_x  + \tfrac{1}{2} q x u \big) , 
\end{align}
giving 
\begin{align}
P_{3{\rm a}} & = a(t)^{-1} u_t + q x u_x  + \tfrac{1}{2} u , 
\label{P.case2a}\\
P_{3{\rm b}} & =  q x a(t)^{-1} u_t + \tfrac{1}{2} (q^2 x^2 + c^2 a(t)^{-2} ) u_x  + \tfrac{1}{2} q x u 
\label{P.case2b}
\end{align}
since $f_{3}(t)=e^{2A}$ from expression \eqref{f3}. 
Hence, the variational symmetries in canonical form are given by 
\begin{align}
\X_{3{\rm a}} & = -q (t+t_1)\partial_t - q x\partial_x  + \tfrac{1}{2} u\partial_u , 
\label{symm.case2a}
\\
\X_{3{\rm b}} & = -q^2 x (t+t_1) \partial_t - \tfrac{1}{2} q^2( x^2 + c^2 (t+t_1)^2 )\partial_x  + \tfrac{1}{2} q x u\partial_u
\label{symm.case2b}
\end{align}
where $a(t)^{-1} = q(t+t_1)$ with $t_1=\tfrac{1}{a_0 q}$ from expression \eqref{a.case2}.
The first symmetry \eqref{symm.case2a} generates 
a dilation on $(t+t_1,x)$ combined with a scaling on $u$: 
\begin{equation}\label{symmgroup.case2a}
t+t_1\to e^{-\epsilon} (t+t_1),
\quad
x\to e^{-\epsilon} x,
\quad
u\to  e^{\epsilon/(2q)} u .
\end{equation}
In contrast, the second symmetry generates a conformal transformation:
\begin{equation}\label{symmgroup.case2b}
t+t_1\to \omega(t+t_1,x) (t+t_1), 
\quad
x\to \omega(t+t_1,x)(x +\tfrac{1}{2}\epsilon(x^2 -c^2(t+t_1)^2)),
\quad
u\to  \omega(t+t_1,x)^{1/(2q)} u 
\end{equation}
where
\begin{equation}
\omega(t,x) = 
\frac{1}{(1+\tfrac{1}{2}\epsilon (x+c t))(1+\tfrac{1}{2}\epsilon(x -c t))} 
\end{equation}
is the conformal factor. 
This transformation preserves the light-cone $x^2 - c^2(t+t_1)^2 =0$. 
Conformal transformations are typically seen for nonlinear wave equations 
only in higher spatial dimensions 
(see e.g. \Ref{Strauss-book,AncIva}). 

The last quantity is the pair of sums of generalized light-cone energies \eqref{C4},
which hold for $a(t)$ and $g(u)$ given by expressions \eqref{case3}, \eqref{a.case3}.
For these two energies, 
\begin{equation} 
\delta C_{4\pm}/\delta u_t = 
e^{\pm a_1q x/c} f_{4}(t) (2 a_1 u_t  + a(t) a_1 u)
\end{equation}
gives
\begin{equation}\label{P.case3}
P_{4\pm} =  e^{\pm a_1q x/c} \big( 2\cosh(a_1q(t +t_1))  u_t  +\sinh(a_1q(t +t_1))(\pm 2 c u_x + a_1 u)\big)
\end{equation}
since $f_{4}(t)=e^{2A} \cosh(a_1q(t +t_1))$ from expression \eqref{f4}. 
This yields the pair of variational symmetries
\begin{equation}
\X_{4\pm} = -e^{\pm a_1q x/c} \big( 2\cosh(a_1q(t +t_1)) \partial_t  +\sinh(a_1q(t +t_1))(\pm 2 c \partial_x - a_1 u\partial_u)\big)
\end{equation}
in canonical form. 
These symmetries have a simpler form in terms of the light-cone variables \eqref{lightcone.vars} with $t$ shifted to  $t+t_1$: 
\begin{equation}\label{symm.case3}
\tfrac{1}{2c}\X_{4\pm} = 
-e^{\pm a_1q \zeta_+/c}\partial_{\zeta_+} - e^{\mp a_1q \zeta_-/c}\partial_{\zeta_-} 
\pm\tfrac{a_1}{4c}(e^{\pm a_1q \zeta_+/c} -e^{\mp a_1q \zeta_-/c}) u\partial_u, 
\quad
\zeta_\pm = c(t+t_1) \pm x , 
\end{equation}
which can be expanded into a sum of generators 
similar to the generator \eqref{symm.case1}. 
It is readily seen that the resulting symmetry transformation group describes
a translation on exponentials of the outgoing and ingoing light-cone variables
\begin{equation}\label{symmgroup.tx.case3}
e^{\mp a_1q\zeta_ +/(4c)} \to e^{\mp a_1q\zeta_ +/(4c)} +\epsilon,
\quad
e^{\pm a_1q\zeta_ -/(4c)} \to e^{\pm a_1q\zeta_ -/(4c)} +\epsilon 
\end{equation}
combined with a $t$-dependent scaling on $u$
\begin{equation}\label{symmgroup.u.case3}
u \to \big( (1+\epsilon e^{\mp a_1q \zeta_-/c})(1 -\epsilon e^{\pm a_1q \zeta_+/c}) \big)^{1/(4q)} u . 
\end{equation}
The invariants consist of 
$e^{\pm a_1q \zeta_-/c}+ e^{\mp a_1q \zeta_+/c}$
and 
$e^{\pm a_1(\zeta_+-\zeta_-)/(4c)}u$. 
Note that the first invariant is equivalent to 
$x\mp \tfrac{c}{a_1q}\ln(\cosh(a_1q(t+t_1)))$ 
which describes a nonlinear light-cone type variable. 
This symmetry is very novel. 

Finally, from Proposition~\ref{prop:waveeqn.ln.nonlin.conslaws},
the nonlinear wave equation \eqref{waveeqn.ln.nonlin} 
with removable damping \eqref{a.sol} 
possesses three conserved quantities \eqref{ln.nonlin.ener}, \eqref{ln.nonlin.mom}, \eqref{ln.nonlin.boostmom}.
The quantity \eqref{ln.nonlin.mom} is a special case of the generalized momentum \eqref{C1} 
which holds regardless of the form of the damping. 
For the quantities \eqref{ln.nonlin.ener} and \eqref{ln.nonlin.boostmom}, 
\begin{equation} 
\delta E/\delta u_t = e^{2A} (u_t +\tfrac{1}{2}a(t) u),
\quad
\delta K/\delta u_t = e^{2A}\big( x (u_t+\tfrac{1}{2}a(t) u)  + c^2 t u_x \big)
\end{equation}
respectively give 
\begin{equation}\label{P.ln.nonlin}
P =  u_t +\tfrac{1}{2}a(t) u,
\quad
P = x (u_t+\tfrac{1}{2}a(t) u)  + c^2 t u_x , 
\end{equation}
which yield the variational symmetries 
\begin{align}
\X & = -\partial_t  +\tfrac{1}{2}a(t) u\partial_u,
\label{symm1.ln.nonlin}\\
\X & = -x\partial_t  -c^2t\partial_x +\tfrac{1}{2}x a(t) u\partial_u
\label{symm2.ln.nonlin}
\end{align}
in canonical form. 
These two symmetries respectively generate the transformation groups
\begin{equation}\label{symmgroup1.ln.nonlin}
t\to t -\epsilon,
\quad
u\to e^{-\tfrac{1}{2}\int^{t-\epsilon}_{t} a(y)\,dy} u, 
\end{equation}
which describes a $t$-translation combined with a $t$-dependent scaling on $u$, 
and 
\begin{equation}\label{symmgroup2.ln.nonlin}
t\to \cosh(c\epsilon) t -  \sinh(c\epsilon) x/c,
\quad
x\to \cosh(c\epsilon) x  -  \sinh(c\epsilon) c t,
\quad
u\to e^{-\tfrac{1}{2}\int^{t-\epsilon}_{t} a(y)\,dy} u, 
\end{equation}
which describes a Lorentz boost on $(t,x)$ combined with a $t$-dependent scaling on $u$.

\section{Concluding remarks}\label{sec:conclude}

All low-order conservation laws have been found for 
a general class of damped nonlinear wave equations \eqref{waveeqn},
as summarized by the classifications stated in 
Proposition~\ref{prop:waveeqn.ln.nonlin.conslaws} in addition to 
Propositions~\ref{prop:lin.multr} and~\ref{prop:nonlin.multr}. 
These classifications can be viewed as an instance of the general inverse problem 
for conservation laws \cite{PopBih}, 
since the results determine all specific wave equations of the general form \eqref{waveeqn}
for which a conserved quantity of first-order in derivatives of $u$ is admitted. 

The admitted conservation laws turn out to be characterized by multipliers 
that are linear in $(u_t,u_x)$, 
which correspond to variational point symmetries. 
In particular, there are no conservation laws with multipliers 
that are nonlinear in $(u_t,u_x)$, and correspondingly, 
no variational (non-point) contact symmetries exist. 

One of the conservation laws describes a generalized momentum,
which holds for arbitrary damping $a(t)$ and nonlinearity $g(u)$. 
The other conservation laws depend essentially on 
the form of both the damping $a(t)$ and the nonlinearity $g(u)$:
\begin{itemize}
\item
generalized Lorentz-boost momentum and generalized energy,
which hold only when $g(u)$ is a log nonlinearity \eqref{g.sol}
and $a(t)$ is removable by a change of variable 
\item
dilational energy and conformal momentum,
which hold only when $g(u)$ is a power nonlinearity plus a mass term \eqref{case1},
and $a(t)$ is a constant 
\item
generalized light-cone energies,
which hold only when $g(u)$ is a power nonlinearity (with no mass term) \eqref{case2},
and $a(t)$ is also a power \eqref{a.case2}
\item 
sums of ingoing and outgoing generalized light-cone energies,
which hold only when $g(u)$ is a power nonlinearity plus a mass term \eqref{case3},
and $a(t)$ is a $\tanh$ \eqref{a.case3}
\end{itemize}

The generalized Lorentz-boost momentum and generalized energy,
as well as the generalized light-cone energies, 
are modifications of conserved quantities known for 
undamped wave equations \eqref{undamped.waveeqn}. 
In contrast, the dilation energy and the conformal momentum 
have no counterpart among those known conserved quantities. 
Instead they are similar to conserved quantities that arise 
for undamped wave equations in higher dimensions,
when $g(u)$ is a certain dimension-dependent power nonlinearity
\cite{Strauss-book,AncIva}. 

All of these new results are relevant for analysis of wave equations \eqref{waveeqn},
since the existence of conserved quantities allows for refinement of theorems
on decay, blow up, and stability.

For future work,
the present results can be naturally extended to 
damped wave equations in higher dimensions \cite{inprogress1} 
and wave equations with nonlinear damping,
and applications in analysis can be expected to emerge.
In particular, 
the symmetries can be used to obtain exact (group-invariant) solutions 
\cite{inprogress2}
whose behaviour may illustrate interesting features such as blow-up and long-time asymptotics.

\section*{Acknowledgements}
SCA is supported by an NSERC Discovery research grant.
APM, TMG, and MLG warmly acknowledge the financial support from the \textit{Junta de Andaluc\'ia} FQM-201 group.

\appendix
\section*{Appendix: Computational remarks}\label{app:computation}

The overdetermined systems leading to 
Proposition~\ref{prop:waveeqn.ln.nonlin.conslaws}
and both Propositions~\ref{prop:lin.multr} and~\ref{prop:nonlin.multr} 
have been solved with use of the software Maple.
In particular, the Maple command `rifsimp' is able to yield a complete classification of 
all cases for which the overdetermined system can be brought to an involutive form 
such that a solution exists,
with $c$ being an arbitrary parameter. 
The solution in each case has been directly verified to satisfy the system. 

\subsection*{Steps in the computation for Proposition~\ref{prop:waveeqn.ln.nonlin.conslaws}}

In this computation, the unknown is $P(t,x,\tilu,\tilu_t,\tilu_x)$. 
Since `rifsimp' is unable to handle non-polynomial variables, 
the function $\tilg(\tilu) = (\sigma -\sigma_0 +\kappa\ln|\tilu|)\tilu$ is replaced 
by an equivalent ODE $(\tilu(\tilg(\tilu)/\tilu)')'=0$. 
It is convenient to divide the computation into two cases: 
(i) $P(t,x,\tilu,\tilu_t,\tilu_x)$ is nonlinear in at least one of  $\tilu_t$, $\tilu_x$;
(ii) $P(t,x,\tilu,\tilu_t,\tilu_x)$ is linear in both $\tilu_t$, $\tilu_x$. 

In case (i), 
the determining equation \eqref{multr.deteqn.ln.nonlin} splits into a system of 6 PDEs. 
Then `rifsimp' is run on this system augmented by the ODE for $\tilg(\tilu)$ 
and the conditions $\tilg(\tilu)\not\equiv0$, 
$P_{\tilu_t\tilu_t}^2 +P_{\tilu_x\tilu_x}^2 + P_{\tilu_t\tilu_x}^2 \not\equiv 0$,
while $c$ is arbitrary. 
The output yields that the system has no solution. 

In case (ii), the first step is substitution of 
$P(t,x,\tilu,\tilu_t,\tilu_x) = P_0(t,x)+ P_1(t,x) u_t + P_2(t,x) u_x$
into the determining equation \eqref{multr.deteqn.ln.nonlin}. 
Splitting again yields a system of 19 PDEs, which are augmented by 
the ODE for $\tilg(\tilu)$ plus the conditions 
$\tilg(\tilu)\not\equiv0$ and $P_0{}^2 +P_1{}^2 + P_2{}^2\not\equiv 0$,
with $c$ being arbitrary. 
The next step is to run `rifsimp'. 
The output consists of a decoupled system of 7 linear PDEs for the $P$s, 
which are easy to integrate. 
The solution is given by the three multipliers \eqref{undamped.multrs}.
This yields Proposition~\ref{prop:waveeqn.ln.nonlin.conslaws}.

\subsection*{Steps in the computation for Proposition~\ref{prop:nonlin.multr}}

This computation for multipliers $Q(t,x,u,u_t,u_x)$ is similar to case (i) 
in the previous computation, except that $g(u)$ and $a(t)$ are also unknowns. 
The determining equation \eqref{nonlin.multr.deteqn} splits into 
a system of 6 PDEs,
and then `rifsimp' is run with the conditions that $c$ is arbitrary
and that 
$Q$ must be nonlinear in $(u_t,u_x)$, 
$g(u)$ must be nonlinear in $u$, 
and $a(t)$ must be non-zero:
$Q_{u_t u_t}^2 +Q_{u_x u_x}^2 + Q_{u_t u_x}^2 \not\equiv 0$,
$g''(u)\not\equiv0$,
$a(t)\not\equiv 0$. 
The output of `rifsimp' yields that the system has no solution. 
Note that the intrinsic damping condition \eqref{cond:intrinsic} is not needed, 
and thus Proposition~\ref{prop:nonlin.multr} holds for the whole class of
damped nonlinear wave equations \eqref{waveeqn}.

\subsection*{Steps in the computation for Proposition~\ref{prop:lin.multr}}

The determining equation \eqref{multr.deteqn.ln.nonlin} 
splits into a system of 19 PDEs whose unknowns are 
$F_1(t,x,u)$, $F_2(t,x,u)$, $F_4(t,x,u)$, in addition to $g(u)$ and $a(t)$.
At least of one the $F$s must be non-zero, namely 
\begin{equation}\label{cond:Fs}
F_1{}^2 + F_2{}^2 + F_4{}^2 \not\equiv 0
\end{equation}
Additionally, $g(u)$ and $a(t)$ must satisfy the conditions \eqref{cond:nonlin.damped}
used in the computation for Proposition~\ref{prop:nonlin.multr},
while $c$ is arbitrary. 

The computation is divided into two three main steps. 
The first step is running `rifsimp' with lowest priority assigned to $g(u)$ 
because this reduces the number of cases in the output. 
Two distinct solution cases are obtained. 

In one case, 
$a(t)$ and $g(u)$ are arbitrary, 
and the $F$s satisfy 5 simple PDEs which are easily solved.
The solution gives the first part of Proposition~\ref{prop:lin.multr}.

In the other case, 
there are 6 PDEs in which both $a(t)$ and $g(u)$ are coupled to the $F$s,
and 5 additional PDEs which involve only the $F$s.
The latter PDEs are straightforward to solve and give 
\begin{equation}\label{Fs}
F_1 = \tfrac{1}{2}\tilde F_1(t)F_2{}_x,
\quad
F_2 = \tilde F_2(t,x),
\quad
F_4 = \tilde F_4(t) + \tfrac{1}{2}\tilde F_1(t)F_2{}_x a u . 
\end{equation}
Substitution back into the first 6 PDEs yields a coupled PDE system 
for $a(t)$, $g(u)$, $\tilde F_1(t)$, $\tilde F_4(t)$, $\tilde F_2(t,x)$. 

For the next step, 
the intrinsic damping condition \eqref{cond:intrinsic} needs to be imposed. 
The form of this condition is not directly usable as an input in `rifsimp' 
because it must hold for all values of the parameter $\mu$. 
There is a straightforward way to reformulate the condition 
by returning to the original ODEs \eqref{g.a.ODEs} for $a(t)$ and $g(u)$,
whose negation yields the condition. 
First, solving for the parameters gives
$\kappa = -(a'(t) + a''(t)/a(t))$ and $\mu = u - g(u)/(g'(u)-\kappa)$. 
Next, differentiating to eliminate the parameters leads to 
\begin{equation}
(a'(t) + a''(t)/a(t))'=0,
\quad
g(u)/( g'(u) +a'(t) + a''(t)/a(t) )=1 . 
\end{equation}
These equations are equivalent to the ODEs \eqref{g.a.ODEs},
and hence their negation provides an alternative formulation of condition \eqref{cond:intrinsic}:
\begin{equation}\label{cond:intrinsic.improved}
\big( (a'(t) + a''(t)/a(t))' \big)^2 + \big( 1 - g(u)/( g'(u) +a'(t) + a''(t)/a(t) ) \big)^2 \not\equiv 0 .
\end{equation}

The final step consists of running `rifsimp' on the coupled system 
for $a(t)$, $g(u)$, $\tilde F_1(t)$, $\tilde F_4(t)$, $\tilde F_2(t,x)$,
along with the conditions \eqref{cond:intrinsic.improved} and \eqref{cond:nonlin.damped}
as well as condition \eqref{cond:Fs} after substitution of expressions \eqref{Fs}. 
The computation simplifies by specifying in the input of `rifsimp' that 
the $\tilde Fs$ have first priority while $g(u)$ has second priority. 
The output yields two cases. 
In both cases, there is a decoupled ODE for $a(t)$; 
an ODE for $g(u)$ in which $a(t)$ appears; 
and 5 PDEs involving all of the unknowns. 
One case contains an additional (`Constraint') ODE for $g(u)$ 
which is nonlinear in its highest-order derivative. 

For both cases, the coupled system of ODEs and PDEs 
requires several non-trivial steps to solve and cannot be done automatically 
by Maple commands `dsolve' and `pdsolve' for reasons that will now be explained. 

For the case with no `Constraint', 
the ODE for $a(t)$ is third order:
$a''' = (aa'' +a'{}^2) a''/(a a')$.
Inspection and elementary integration shows that its general solution 
has 3 different branches:
(1) $a''=0$; 
(2) $a' = -q a^2$, $q=\const\neq 0$; 
(3) $a' = q (a_1{}^2 -a^2)$, $q,a_1=\const\neq 0$.
Each of these ODEs is simple to solve. 
Note, however, it is not possible to obtain all of these branches automatically 
if `dsolve' is used, since in general `dsolve' yields 
only the generic branch when solving a given ODE. 
Consequently, the rest of the computation must be split by hand into 3 subcases,
corresponding to the 3 distinct solution branches. 

Branch (1) has $a=a_0(t+t_1)$, $a_0,t_1=\const$. 
The ODE for $g(u)$ reduces to $g''= a_1(a_1+g')/g$, which is straightforward to solve
as it is invariant under translation in $u$ and scaling of $u$ and $g$. 
But, after substitution of $a$ and $g$, 
the intrinsic damping condition \eqref{cond:intrinsic.improved} turns not to be satisfied. 
Hence this subcase does not have a solution. 

Branch (2) has $a= 1/(q(t+t_1))$, $t_1=\const$,
and the ODE for $g(u)$ is $g'' = \tfrac{2q}{1+2q} g'{}^2/g$, 
which is separable and yields $g= k(u+u_0)^{1+2q}$, $u_0=\const$. 
The intrinsic damping condition \eqref{cond:intrinsic.improved} is satisfied. 
Substitution of $a$ and $g$ into the remaining 5 PDEs 
yields a system of simple linear DEs for $\tilde F_1(t)$, $\tilde F_4(t)$, $\tilde F_2(t,x)$.
This system is straightforwardly solved by an iterative process of 
integrating the elementary DEs first, 
substituting back into the remaining DEs, 
and separating with respect to $t$ and $x$. 
Alternatively, the system of 5 equations can be solved at once 
by using the Maple command `pdsolve'. 
Since this command in general yields only the generic solution, like `dsolve', 
it is necessary to check by inspection that no solution branches have been missed. 
The result gives case (ii) of Proposition~\ref{prop:lin.multr}.

Branch (3) has $a=a_1\tanh(\tfrac{q a_1}{2}(t+t_1))$, $t_1=\const$.
The ODE for $g(u)$ is $g'' = \tfrac{q}{4(1+2q)} (2g' + q a_1{}^2)(4g' - a_1{}^2)/g$, 
which is invariant under translation in $u$ and scaling of $u$ and $g$. 
This leads to $g = k(u+u_0)^{1+2q} + \tfrac{a_1{}^2}{4}(u+u_0)$. 
Again the intrinsic damping condition \eqref{cond:intrinsic.improved} is satisfied. 
For the remaining 5 PDEs, 
the previous integration process can be used, 
but this leads to term swell due to substitution of $a$, 
and requires a lot of intermediate simplifications. 
Even worse, running `pdsolve' produces a lengthy output 
which needs considerable simplification and further requires a difficult manual check 
that no solution branches have been missed. 
Instead, a more efficient process can be used where $a$ is not substituted
into the system of 5 PDEs. 
Integration of the simplest PDEs with respect to $t$ derivatives is done first, 
which involves introducing an auxiliary function $B(t) = e^{\int (a(t)\,dt +\frac{1}{2}qa_1{}^2\int 1/a(t))\,dt}$. 
Then the separation step with respect to $t$ and $x$ is carried out 
by using the differential relation $B'(t)/B(t) = a(t) +\frac{1}{2}qa_1{}^2 /a(t)$. 
The final result gives case (iii) of Proposition~\ref{prop:lin.multr}.

This completes the case with no `Constraint'. 
The remaining case with a `Constraint' is solved as follows. 

The ODE for $a(t)$ is elementary: $a'=0$, whence $a=a_0=\const$.  
However, the two ODEs for $g(u)$ are quite complicated:
one is 4th order; 
the other one, which is given by the `Constraint', is lower order 
but nonlinear in its highest-order derivative terms. 
Rather than seeking all possible solution branches by hand, or relying on `dsolve', 
it is possible to obtain a simpler ODE by examining the PDEs 
in which $g(u)$ is coupled to $\tilde F_1(t)$, $\tilde F_4(t)$, $\tilde F_2(t,x)$. 
Three of these PDEs can be separated with respect to $u$, 
yielding three different ODEs for $g(u)$. 
Their consistency is ensured by the `Constraint' ODE. 
One of these ODEs is separable and can be integrated straightforwardly to obtain
$g = k(u+u_0)^{1+p} + \tfrac{a_0{}^2}{4}(u+u_0) +g_0$, $g_0=\const$. 
Substitution into the `Constraint' ODE gives $g_0=0$. 
The remaining steps are similar to branch (2) of the previous case,
and the resulting solution gives case (i) of Proposition~\ref{prop:lin.multr}.

\end{document}